\documentclass[numbers,authoryear, 12pt]{article}
\usepackage[T1]{fontenc}
\usepackage[utf8]{inputenc}
\usepackage[top=3cm, bottom=3cm, left=3cm, right=3cm]{geometry}
\usepackage{exscale}
\usepackage{amssymb}
\usepackage{dsfont}
\usepackage{mathpazo}
\usepackage{amsmath, amsthm, mathtools, amsfonts}
\usepackage{tikz}
\usetikzlibrary{positioning}
\usepackage{fullpage}
\usepackage{float}
\usepackage{subfig}
\usepackage{threeparttable}
\usepackage{rotating}
\usepackage{enumerate}   
\usepackage{tabularx}
\DeclareFontFamily{U}{mathx}{}
\DeclareFontShape{U}{mathx}{m}{n}{<-> mathx10}{}
\DeclareSymbolFont{mathx}{U}{mathx}{m}{n}
\DeclareMathAccent{\widehat}{0}{mathx}{"70}
\DeclareMathAccent{\widecheck}{0}{mathx}{"71}
\usepackage{caption}
\usepackage{booktabs}
\usepackage{eurosym}
\usepackage{array,multirow,comment}
\newcolumntype{H}{>{\leftrightarrowalse}c<{\fi}@{}}
\usepackage[normalem]{ulem}

\usepackage{pdflscape}
\usepackage[en-US]{datetime2}
\makeatletter
\newcommand{\vast}{\bBigg@{5}}
\newcommand{\Vast}{\bBigg@{6}}
\makeatother

\usepackage{setspace}
\def\spacingset#1{\renewcommand{\baselinestretch}%
{#1}\small\normalsize} \spacingset{1}
\usepackage{authblk}

\usepackage[round,colon,authoryear,sort&compress]{natbib}
\usepackage{hyperref}
\hypersetup{colorlinks=true, citecolor=blue, filecolor=green, linkcolor=blue}

\newtheorem{lemma}{Lemma}
\newtheorem{proposition}{Proposition}
\newtheorem{definition}{Definition}
\newtheorem{assumption}{Assumption}
\newtheorem{corollary}{Corollary}
\providecommand{\keywords}[1]
{
 \small	
 \textbf{\textit{Keywords---}} #1
}
\providecommand{\JEL}[1]
{
 \small	
 \textbf{\textit{JEL Classification---}} #1
}
\title{\Large \textbf{Heterogeneity in peer effects for binary outcomes}}

\vspace{1cm}
\author[1,*]{Mathieu Lambotte}

\affil[1]{Univ Rennes, CNRS, CREM – UMR6211, F-35000 Rennes France}

\affil[*]{\small mathieu.lambotte@univ-rennes.fr \\~\

Please click \href{https://drive.google.com/file/d/1N-_nWZngUvc94_n4p_DPaLwNKm1fdWov/view?usp=sharing}{here} to ensure that you are reading the last version.}

	\date{May 21, 2026}

\begin{document}

\maketitle

\begin{abstract}

\noindent 
I introduce heterogeneity into the analysis of peer effects arising from conformity by allowing peer-effect parameters to vary across agents’ actions. Using a structural model based on a simultaneous network game with incomplete information, I derive conditions that guarantee the uniqueness of the equilibrium and the identification of heterogeneous peer-effect parameters.
Applying the model to data on smoking and alcohol consumption among secondary school students, and conducting policy simulations, I show that assuming a homogeneous preference for conformity leads to biased estimates of peer effects and ex ante policy evaluations. \\

 \\

\JEL{C31, D85, I12} \\

\begin{center} \small \textbf{Acknowledgments}\end{center}

\noindent A previous version of this paper, titled ``Identification and estimation of asymmetries in peer effects for binary outcomes'' has been circulating since May 2024. I thank Philip Ushchev, Bernard Salanié, Francis Bloch and Aristide Houndetoungan for their helpful comments and suggestions. I am also grateful to Jordan Norris and other participants of the 2025 World Congress of the Econometric Society, and to Josselin Thuillez, who financed the access to the Add Health data through his chair "Health Economics" from Rennes Metropole.
This research uses data from Add Health, a program directed by Kathleen Mullan Harris and designed by J. Richard Udry, Peter S. Bearman, and Kathleen Mullan Harris at the University of North Carolina at Chapel Hill, and funded by Grant P01-HD31921 from the Eunice Kennedy Shriver National Institute of Child Health and Human Development, with cooperative funding from 23 other federal agencies and foundations. Special acknowledgment is given to Ronald R. Rindfuss and Barbara Entwisle for assistance in the original design. Information on how to obtain Add Health data files is available on the Add Health website (http://www.cpc.unc.edu/Add Health). No direct support was received from Grant P01-HD31921 for this research.

\end{abstract}
\clearpage

\spacingset{1.5}
\section{Introduction}
With the growing recognition that economic behaviors are embedded in social interactions and the increasing availability of network data, the identification and estimation of peer effects have generated a rich theoretical and empirical literature \citep[for recent reviews, see][]{zenou2025,Bramoulle2019}. Theoretical models of social interactions typically allow for two-way heterogeneous peer effect coefficients \citep[e.g.,][]{ballester2006}, meaning that the influence of individual $j$ on $i$ may differ from her influence on $k$, or from $k$’s influence on $i$. While such flexibility is conceptually appealing, econometric implementation is untractable as identifying $n(n-1)$ parameters from $n$ behavioral equations is infeasible. Consequently, most empirical work has relied on the more tractable linear-in-means model, in which agents react to the average behavior of their peers, and peer effects are constrained to be homogeneous. \\

This paper focuses on binary outcomes and the peer effects that arise from preferences for conformity, a setting in which individuals incur a cost when deviating from the average behavior of their friends. In this context, the assumption of a homogeneous conformity parameter is particularly restrictive. Consider high school students' decisions about smoking. Under homogeneity, smoking when 70\% of friends do \textit{not} smoke is assumed to generate the same utility loss as \textit{not} smoking when 70\% of friends do smoke. Yet this symmetry may be unrealistic: smokers may care little about the norm, due to addiction or a rebellious self-image, whereas non-smokers may face strong pressure to conform, as deviating from the majority behavior could lead to ostracization.

Imposing homogeneous peer effects can thus misrepresent preferences and lead to misguided policy recommendations. Suppose a social planner wishes to shift behavior toward a \textit{lower} equilibrium.\footnote{Henceforth, a \textit{lower} (\textit{higher}) equilibrium refers to a game's outcome in which more players select the \textit{low} (\textit{high}) action.} She could finance campaigns that strengthen the taste for conformity when choosing the high action or that weaken the taste for conformity when choosing the low action. Examples of such policies are anti-smoking campaigns stressing the stigma of smoking (e.g., the U.S. FDA’s The Real Cost campaign) or campaigns highlighting that not drinking does not hinder social participation (e.g., “You don’t need alcohol to have fun”). For instance, an increase in the preference for conformity when smoking would decrease the utility of smoking for agents whose social norm is \textit{not} to smoke. Conversely, a decrease in the preference for conformity when \textit{not} drinking alcohol would increase the utility of \textit{not} drinking for agents whose social norm is to drink. With homogeneous conformity parameters, however, the planner can act only unidimensionally, either strengthening or weakening the pressure to conform. These policies may be counterproductive: for instance, reducing the preference for conformity unidimensionally may induce some non-smokers to start smoking. 

Action-specific peer effects can also explain why social nudges sometimes backfire \citep{schultz2007}. If the taste for conformity when choosing the high action is null, a global-norm campaign such as the 2014 “Finish It’’ message—“Today, only 9\% of teens smoke’’—might inadvertently lead to a \textit{higher} equilibrium. In this scenario, smokers do not care about their distance to the social norm and thus face no additional penalty after learning that smoking is rare. Conversely, non-smokers' utility may be penalized if the preference for conformity when choosing the low action is positive. By contrast, if the preference for conformity is homogeneous and positive, such nudges would reliably reduce smoking prevalence. \\

The main contribution of this paper is to show that action-specific peer effects can be identified and estimated for binary outcomes. More precisely, it is possible to separately identify $(i)$ the taste for conformity when the agent chooses the low action, and $(ii)$ the taste for conformity when she chooses the high action. This allows researchers to directly infer whether conformity pressures are stronger for the low or high action. This result relies on allowing the social-distance function, which measures the distance between the agent’s action and the average action of her friends, to vary depending on the action chosen by the agent. This generalizes the standard quadratic specification used in the literature \citep[see e.g.,][]{akerlof1997,bernheim1984,brock_discrete_2001}, which assumes that conformity pressure is homogeneous across actions. Hence, the model with an action-specific preference for conformity effects nests the homogeneous specification, enabling likelihood-ratio tests of whether conformity is action-specific. In addition, when heterogeneity is introduced in the tastes for conformity, I uncover that players' best responses depend nonlinearly on the expected social norm and are also influenced by the variance of the norm. Because of this additional flexibility, the model can generate multiple equilibria even under parameter configurations that would yield a unique equilibrium in the homogeneous case. Estimates from homogeneous conformity models may therefore provide a misleading picture when action-specific tastes for conformity are present, both by imposing homogeneity in conformity and by failing to account for potential equilibrium multiplicity that is relevant for counterfactual and policy analysis. \\

This paper complements recent work that has begun to introduce heterogeneity 
into empirical peer effect models. One strand of the literature conditions 
heterogeneity on observable characteristics such as demographics 
\citep[e.g.,][]{nakajima2007,hsieh2018,houndetoungan_count_2020} or network 
centrality \citep{lin2017}, allowing researchers to examine, for example, 
whether peer effects are stronger among individuals of the same gender or 
increase with the relative centrality of friends.

A second strand studies heterogeneity in the preference for conformity that stems from the definition of the norm. For continuous outcomes, \cite{boucher2023} allow the norm to depend on the minimum or maximum friend outcome rather than on the mean, while \cite{houndetoungan2025} allows peer effects to vary across quantiles of friends' outcomes. 

A third approach, to which this paper belongs, introduces action-specific peer effects. \cite{Badev2021} develops a spillover model for binary outcomes in which agents choose both their actions and their friends within a dynamic framework under complete information. \cite{xu2018} proposes a spillover model in an incomplete information setting, where peer effect parameters vary across combinations of an agent's action and her friends' actions. However, for identification, the author imposes a normalization that sets all peer effect parameters to zero when either the agent or her friends choose the low action, so that this approach only recovers the effect of friends choosing the high action on the utility of choosing the high action. The present paper is able to separately identify conformity pressures for both low and high actions.

Spillover models, such as those of \cite{xu2018} and \cite{Badev2021}, assume that actions are strategic complements, so that the utility of choosing a given action increases with the share of friends choosing that same action. \cite{brock_discrete_2001} show that 
spillover and conformity models generate the same best-response function 
under homogeneous peer effects. I show that this equivalence breaks down 
in the presence of action-specific heterogeneity.\footnote{See Online 
Appendix A.1 and A.2 for a proof.} The two approaches therefore generate 
distinct best responses, and I show how a standard Wald test can determine 
whether the data are consistent with the spillover model, the conformity 
model, or neither, when action-specific heterogeneity and isolated agents 
are present in the data.\footnote{The specification test is presented in 
Online Appendix A.3.} The continuous-outcome model of \cite{boucher2023} 
also proposes a specification test to distinguish between spillover and 
conformity microfoundations, although it does not require action-specific 
heterogeneity in peer effects. Identifying the correct microfoundation 
matters beyond estimation: conformity and spillover models have distinct 
welfare implications and motivate different policy interventions 
\citep{brock_discrete_2001,ushchev2020}. \\

Section \ref{section1} presents the microfoundations for the conformity model as a simultaneous game on a network with incomplete information, in which players do not observe their friends' behaviors but form rational expectations based on the available public information. 
In Section \ref{section2}, I derive a sufficient condition for the existence of a unique Bayes-Nash equilibrium (BNE) and propose its comparative statics analysis.
Section \ref{section3} presents the conditions for identification of the action-specific conformity parameters and outlines the structural estimation strategy. 
Finally, I apply the model to the data in Section \ref{section4} and provide empirical evidence of heterogeneity in conformity preferences using the Add Health data, which includes network and individual-level data on secondary school students from 126 U.S. schools. This analysis primarily serves as a proof of concept and revisits an important paper by \cite{lee_binary_2014}, which develops a network-based estimation strategy for the homogeneous peer effects model proposed by \cite{brock_discrete_2001}, focusing on students' smoking behavior. I extend the analysis to alcohol consumption, illustrating how the magnitude of the action-specific heterogeneity varies across behaviors. 
The empirical results show that imposing a homogeneous, rather than heterogeneous, preference for conformity leads to misleading estimates of peer effects for smoking, but not for drinking. In addition, for smoking, I show that the data are not consistent with a pure spillover model and I cannot reject the null hypothesis that they are consistent with a pure conformity model at a significance level below 0.2\%. Finally, I conduct policy simulations to assess the extent to which heterogeneity in conformity preferences impacts the effectiveness of social marketing campaigns and social nudges.

\section{Microfoundations} \label{section1}
This section presents the microfoundations of the model, which are based on a simultaneous game of incomplete information with a set of $n$ players, indexed by $i$ and denoted by $\mathcal{N}=\{1,\dots,n\}$.
\subsection{Network Game with Incomplete Information}
 \textbf{Outcome}. I focus on a binary outcome, $y_i \in \mathcal{Y}_i$, which indicates the behavior or \textit{action} of player $i \in \mathcal{N}$. I define $\mathcal{Y}_i=\left\{0,1\right\}$, where a value of one is given to the \textit{high} action (e.g., smoking, exerting effort, or adopting an innovation) and a value of zero to the \textit{low} action. \\%The action profile of the $n$ players, the outcome of the game, is given by $\mathbf{y}=(y_1,\dots,y_n)^{\prime}$, where $y_i \in \mathcal{Y}_i$ for each $i\in \mathcal{N}$. 
 
\noindent \textbf{Players' Characteristics}. Players' characteristics are described by an idiosyncratic variable $\alpha_i$, which is observed by all players. In the empirical application, $\alpha_i$ depends on students' characteristics, the average characteristics of their friends, and school fixed effects. Players are also equipped with private preference shocks $\{\epsilon_i(y_i)\}_{y_i\in \mathcal{Y}_i}$ for both the high and low actions, which are unknown to the other players (and the researcher).  \\

\noindent \textbf{Network}. Players interact on a network, which is represented as an $n\times n$ binary matrix $\mathbf{A}=[a_{ij}]$, where \(a_{ij}=1\) if players \textit{i} and \textit{j} are friends, and 0 otherwise. Self-influence is not allowed, i.e., $a_{ii}=0 \; \forall i \in \mathcal{N}$ and links might be directed or undirected. %Additionally, \(a_{ij}\) may differ from \(a_{ji}\), reflecting the directed nature of the network, where player \textit{i} may consider player \textit{j} a friend, but player $j$ may not reciprocate. 
 I denote players' number of friends by $d_i=\sum\limits_{j\neq i}a_{ij}$ and define $\mathbf{a}_i$ as the $i^{th}$ row of $\mathbf{A}$.\footnote{Henceforth, for any matrix $\mathbf{X}$, the vector $\mathbf{x}_k$ denotes its $k^{th}$ row.} I also use a row-normalized matrix $\mathbf{G}=[g_{ij}]$, which is obtained by row-normalizing $\mathbf{A}$, i.e., $g_{ij}=\frac{a_{ij}}{d_i}$. \\

\noindent \textbf{Preferences}. Following \cite{brock_discrete_2001}, I specify player $i$'s utility function $U(\cdot)$ as an additive function of three components:
\begin{equation}\label{eq1}
 U_i(y_i,\mathbf{y}_{-i})=\alpha_i y_i-\mathds{1}\{d_i>0\}S^{quad}_i(y_i,\mathbf{y}_{-i})+\epsilon_i(y_i)
\end{equation}
where $\mathds{1}\{\cdot\}$ is the indicator function, such that $\mathds{1}\{d_i>0\}$ takes the value one if player $i$ is not isolated and zero if she is isolated. To simplify the exposition of the model, I assume hereafter that there are no isolated players. In the empirical application, however, I account for the fact that the utility function simplifies to $U_i(y_i,\mathbf{y}_{-i})=\alpha_i y_i+\epsilon_i(y_i)$ for isolated players. Note that the specification of the utility function in Equation \eqref{eq1} ensures that $\alpha_i$ is normalized to zero for $y_i=0$. This normalization does not affect the equilibrium's action profile $\mathbf{y}$ but is necessary for identifying the parameters associated with $\alpha_i$.\\

\noindent  \textbf{Homogeneous Social Distance Function}. $S^{quad}(y_i,\mathbf{y}_{-i})$ is the quadratic social distance function, introduced by \cite{bernheim1984} and \cite{akerlof1997}, widely used in network games with binary or continuous action spaces \cite[see e.g.,][]{brock_discrete_2001,blume_linear_2015,boucher2023,houndetoungan_count_2020,patacchini2014,ushchev2020,li2009}, and given by:\footnote{A different specification is proposed in \cite{Boucher2016}, where the distance function is defined as $\frac{\beta}{2}\sum\limits_{j\neq i}\left[a_{ij}(y_i-{y_{j}})^2\right]$. In this formulation, the cost of deviating from friends’ behavior scales with the number of friends a player has. I show in Online Appendix B.1 that action-specific heterogeneity in the preference for conformity can also be identified when the social distance function computes the weighted sum of deviations from friends’ actions, rather than the deviation from the average of friends’ actions. However, identification in that case relies on the existence of isolated players.}
\begin{equation}\label{quad}
        S^{quad}_i(y_i,\mathbf{y}_{-i})=\frac{\beta}{2}\left(y_i-\bar{y}_i\right)^2\\
\end{equation}
where $\bar{y}_i=\sum_{j\neq i}g_{ij}y_j$ is the social norm faced by agent $i$. Social distance thus depends on friends' actions and incorporates the game's strategic dimension, the so-called \textit{endogenous peer effects}. 
Because the distance function is quadratic, the social distance increases with the deviation from the norm when $\beta>0$, i.e., $\frac{\partial S^{quad}_i(y_i,\mathbf{y}_{-i})}{\partial (y_i-\bar{y_i})}=\beta(y_i-\bar{y_i})$. This increasing marginal social distance implies that small deviations from the norm have a relatively minor effect on a player’s utility, whereas larger deviations are increasingly costly. Although this feature of the social distance function seems realistic, I show in Online Appendix B.2 that a linear social distance function can also be used, although the presence of isolated players is necessary for identification in that case. The taste for conformity $\beta$ is unrestricted here and may be negative if players are anticonformist and prefer to deviate from the norm, capturing, for instance, status concerns \citep{immorlica2017,fershtman1998} or competitive preferences \citep{azmat2010,lopez2021}. In addition, define $\mathbf{1}_{n-1}$ as the $(n-1) \times 1$ vector of ones and observe that $S^{quad}(0,\mathbf{y}_{-i})=S^{quad}(1,\mathbf{1}_{n-1}-\mathbf{y}_{-i})$ and the distance function is thus \textit{symmetric}. Hence, under the homogeneous specification $S^{quad}_i(\cdot)$, players face the same penalty whether they choose the high or the low action, for a given distance to the norm. \\

\noindent  \textbf{Heterogeneous Social Distance Function}. I propose an alternative heterogeneous specification of the social distance function, $S_i^{het}(\cdot)$, that nests the homogeneous one:
\begin{equation}\label{S}
 S^{het}_i(y_i,\mathbf{y}_{-i})=y_i\left(\frac{\beta^h}{2}\left(y_i-\bar{y}_i\right)^2\right)+(1-y_i)\left(\frac{\beta^l}{2}\left(y_i-\bar{y}_i\right)^2\right)
\end{equation}
 $\beta^h$ captures the taste for conformity when choosing the high action. With binary outcomes, selecting $y_i=1$ implies that $y_i\geq \bar{y}_i$ and thus that the agent's action is (weakly) \textit{above} the norm. $\beta^h$ can thus be interpreted as the cost of being \textit{above} the norm. Conversely, $\beta^l$ measures the taste for conformity when choosing the low action and thus captures the cost of being \textit{below} the norm. $\beta^h$ and $\beta^l$ do not need to have the same sign, and may be negative. For instance, in the case of smoking among teenagers, one might expect $\beta^l>\beta^h \geq0$ if students perceive smoking as socially desirable. In this scenario, students are strongly penalized for not smoking when smoking is the norm among their friends, but they experience a weaker, or even no, penalty for smoking, even if it goes against the norm. In contrast, polluting, a socially undesirable behavior, may be characterized by $\beta^h>\beta^l\geq0$. In that case, agents are strongly penalized for polluting when the norm among their friends is to abstain, but face little or no penalty for not polluting, even if most of their friends pollute.
In general, one may expect the relationship between $\beta^h$ and $\beta^l$ to depend on the social desirability of the behavior under scrutiny. In the limit, $\beta^h \to \infty$ for \textit{prohibited} behaviors, while $\beta^l \to \infty$ for \textit{mandatory} behaviors. Note that the heterogeneous social distance function is asymmetric when $\beta^h\neq\beta^l$, in the sense that $S^{het}(0,\mathbf{y}_{-i})\neq S^{het}(1,\mathbf{1}_{n-1}-\mathbf{y}_{-i})$. \\
%Finally, it is possible to show that the homogeneous taste for conformity, $\beta$, is related to $\beta^h$ and $\beta^l$.

%\begin{proposition}\label{P0}
%$\beta$ is a weighted sum of $\beta^h$ and $\beta^l$, given by $\beta=(1-\lambda)\beta^h + \lambda \beta^l$.
%\end{proposition}
%The proof is provided in Appendix \ref{AppendixF}. If $\lambda \in [0,1]$, then $\beta$ is a weighted average of $\beta^h$ and $\beta^l$, ensuring that $\beta$ lies between $\beta^h$ and $\beta^l$. In the empirical application, we can estimate $\lambda$ as $\hat{\lambda}=\frac{\hat{\beta}-\hat{\beta}^h}{\hat{\beta}^l-\hat{\beta}^h}$. \\

\noindent \textbf{Players' type}. I assume that the distributions of $\{\epsilon_i(y_i)\}_{y_i\in \mathcal{Y}_i}$ are common knowledge and identical for all players. Players are thus facing uncertainty about the realization of their friends' private preference shocks. Define players' type as $\eta_i=\epsilon_i(0)-\epsilon_i(1)$, such that $\eta_i$ corresponds to the private relative preference of player $i$ for the low action over the high one.

\begin{assumption}\label{H1}
\begin{enumerate}[$(i)$]
\item The random variables $\epsilon_i(y_i)$ are distributed iid across actions $y_i\in \mathcal{Y}_i$ and players. 
\item The density $f_\epsilon$ is continuous, integrable and positive everywhere on \(\mathbb{R}\). 
\item $\{\epsilon_i(y_i)\}_{y_i\in \mathcal{Y}_i}$ is independent of $\boldsymbol \alpha$ and $\mathbf{G}$. 
\end{enumerate}
\end{assumption}
Assumption \ref{H1}$(i)$ and the continuity and integrability of $f_\epsilon$ in 
Assumption \ref{H1}$(ii)$ guarantee that $\eta_i=\epsilon_i(0)-\epsilon_i(1)$ admits 
a continuous density $f_{\eta}$. Moreover, the 
strict positivity of $f_\epsilon$ ensures that $f_\eta$ is strictly positive everywhere 
on $\mathbb{R}$, which in turn guarantees that $F_\eta$ is strictly increasing. 
Assumption \ref{H1}$(iii)$ implies that players' actions can be correlated only through 
the network and the characteristics $\boldsymbol{\alpha}$, and that knowing their own type does not provide 
any information on the types, and therefore actions, of other players. To rule out 
endogeneity issues in the econometric model, I further assume that $\mathbf{G}$ and 
$\boldsymbol{\alpha}$ are exogenous, although \cite{Badev2021} and \cite{lambotte2024} 
propose methods to account for network endogeneity in peer effects models with binary 
outcomes. \\

\noindent \textbf{Rational Expectations}. Players are thus uncertain about other players' actions $\mathbf{y}_{-i}$ and do not maximize the utility function in Equation (\ref{eq1}) but its expectation with respect to their beliefs about other players' actions. I assume that players form rational expectations, i.e., \textit{objective} mathematical expectations about other players' actions given the information set available to them. The information set of a given player is $\{\boldsymbol \alpha,\mathbf G,\eta_i\}$. Under Assumption \ref{H1}$(iii)$, players' types are independent of $\mathbf G$ and $\boldsymbol \alpha$, such that, by mutual independence, the information set used to form expectations about other players' actions reduces to $\{\boldsymbol \alpha,\mathbf G\}$. 
\begin{assumption}\label{H2}
Players form rational expectations about other players' actions, conditional on the public information set $\{\boldsymbol \alpha,\mathbf G\}$, denoted as $p_j=\mathbb{E}\left[y_j\mid \boldsymbol \alpha,\mathbf G\right]$.
\end{assumption}
Note that the rational expectations are heterogeneous in the sense that players with different characteristics and positions in the network are expected to play different actions. I denote the rational expectation profile by $\mathbf p=(p_1,\dots,p_n)^{\prime}$ and the expected norm faced by player $i$ as $\bar{p}_i=\sum_{j \neq i}g_{ij}p_j$. \\

\noindent \textbf{Expected utility}. The utility function with the heterogeneous social distance function can be written as:
\begin{equation}\label{U}
 U_i(y_i,\mathbf y_{-i})=\alpha_iy_i-\left( y_i\frac{\beta^h}{2}+(1-y_i)\frac{\beta^l}{2}\right)\left(y_i-\bar{y}_i\right)^2+\epsilon_i(y_i)
\end{equation}
Since only $\mathbf{y}_{-i}$ is unobserved by player $i$, her expected utility is obtained by taking expectations with respect to $\mathbf{y}_{-i}$ and is given by:
\begin{align*}
\label{Ue}
 \mathbb{E}[U_i(y_i,\mathbf{y}_{-i})\mid \boldsymbol \alpha,\mathbf G]&=
\begin{cases} 
\alpha_i- \frac{\beta^h}{2}\left(1-2\bar{p}_i+\mathbf{g}_i\boldsymbol\Sigma  \mathbf{g}^{\prime}_i \right)+\epsilon_i(1) & \text{if } y_i=1 \\
- \frac{\beta^l}{2}\mathbf{g}_i\boldsymbol\Sigma \mathbf{g}^{\prime}_i+\epsilon_i(0) & \text{if } y_i=0 \\
\end{cases}
\end{align*}
where $\bar{p}_i=\sum_{j\neq i}g_{ij}p_j$ and $\mathbf{g}_i\boldsymbol\Sigma \mathbf{g}^{\prime}_i=\mathbb{E}[\bar{y}_i^2\mid \boldsymbol \alpha,\mathbf G]=\mathbb{E}[\bar{y}_i\mid \boldsymbol \alpha,\mathbf G]^2 +\mathbb{V}[\bar y_i\mid \boldsymbol \alpha,\mathbf G]=\bar{p}_i^2+\sum_{j}g_{ij}^2p_j(1-p_j)$, where the last equality holds because $y_j$ is a binary variable and because Assumption \ref{H2} imposes independence of $\{y_i\}_{i=1}^n$ conditional on the public information set $\{\boldsymbol \alpha, \mathbf G\}$.\footnote{To see this, let $\circ$ denote the Hadamard product. Define the operator $\operatorname{diag}(\mathbf a)$, with $\mathbf a$ an $n\times 1$ vector, as an $n\times n$ matrix whose diagonal entries are the elements of $\mathbf a$ and off-diagonal elements are zero, while $\operatorname{diag}(\mathbf A)$, with $\mathbf A$ an $n\times n$ matrix, corresponds to an $n\times 1$ vector that returns the diagonal elements of $\mathbf A$. Then, \begin{align*}
 \mathbb{E}\left[\bar{y}_i^2\mid \boldsymbol \alpha,\mathbf G\right]&=\mathbf{g}_i\mathbb{E}\left[\mathbf y\mathbf y^{\prime}\mid \boldsymbol \alpha,\mathbf G\right]\mathbf{g}_i^{\prime}\\
    &=\mathbf{g}_i\begin{pmatrix}
\mathbb{E}[y_1 \mid \boldsymbol\alpha,\mathbf G] & \mathbb{E}[y_1 y_2 \mid \boldsymbol\alpha,\mathbf G] & \cdots & \mathbb{E}[y_1 y_n \mid \boldsymbol\alpha,\mathbf G] \\
\mathbb{E}[y_1 y_2 \mid \boldsymbol\alpha,\mathbf G] & \mathbb{E}[y_2 \mid \boldsymbol\alpha,\mathbf G] & \cdots & \mathbb{E}[y_2 y_n \mid \boldsymbol\alpha,\mathbf G] \\
\vdots  & \vdots & \ddots & \vdots \\
\mathbb{E}[y_1 y_n \mid \boldsymbol\alpha,\mathbf G] & \mathbb{E}[y_2 y_n \mid \boldsymbol\alpha,\mathbf G] & \cdots & \mathbb{E}[y_n \mid \boldsymbol\alpha,\mathbf G]
\end{pmatrix}\mathbf{g}^{\prime}_i\\
&=\mathbf{g}_i\begin{pmatrix}
p_1 & p_1p_2 & \cdots & p_1p_n \\
p_1p_2 & p_2& \cdots & p_2p_n \\
\vdots  & \vdots& \ddots & \vdots \\
p_1p_n& p_2p_n&\cdots & p_n
\end{pmatrix}\mathbf{g}^{\prime}_i=\mathbf{g}_i\left(\mathbf{p} \mathbf p^{\prime}+\operatorname{diag}(\mathbf p\circ(\mathbf{1}_n-\mathbf p))\right)\mathbf{g}^{\prime}_i=\mathbf{g}_i\boldsymbol\Sigma \mathbf{g}^{\prime}_i,
\end{align*} where the second equality uses the fact that $\mathbb{E}[y_i^2]=\mathbb{E}[y_i]$ because $y_i$ is binary and the third equality holds because $y_i$ and $y_j$ are independent conditional on $\{\boldsymbol\alpha,\mathbf G\}$ under Assumption \ref{H2}.} 
 % and $\operatorname{diag}(\mathbf p\circ(\mathbf{1}_n-\mathbf p))=\begin{pmatrix}
%p_1(1-p_1) & 0 & \cdots & 0 \\
%0 & p_2(1-p_2)& \cdots & 0 \\
%\vdots  &\vdots & \ddots & \vdots \\
%0& 0&\cdots & p_n(1-p_n)
%\end{pmatrix}$As a result, $\mathbb{E}[U_i(y_i=0,\mathbf{y}_{-i})]$ is normalized to zero, except for the part of the utility function that accounts for the social distance. Indeed, if $\beta^l>0$ and $\bar{p}_i\neq0$, non-isolated players bear the cost, proportional to $\mathbf{g}_i\boldsymbol\Sigma  \mathbf{g}^{\prime}_i$, of not conforming to the norm, as $y_i=0\implies y_i<\bar{p}_i$ here. 

\subsection{Best-Response Function}
\textbf{Decision Rule}. Define $\mathbb{E}[\Delta _iU_i]=\mathbb{E}\left[U_i(y_i=1,\mathbf{y}_{-i})- U_i(y_i=0,\mathbf{y}_{-i})\mid \boldsymbol \alpha,\mathbf G\right]$ as the marginal expected utility from playing $y_i=1$ instead of $y_i=0$, such that player $i$ selects the high action only if $\mathbb{E}[\Delta _iU_i] >0$:
\begin{equation}\label{deltaU}
\{y_i=1\} \iff \left\{\alpha_i+\beta^h\left(\bar{p}_i-\frac{1}{2}\right)+ \frac{\Delta \beta}{2}\bar p_i^2 + \frac{\Delta \beta}{2}  \sum_{j\neq i}g_{ij}^2p_j(1-p_j)-\eta_i>0\right\}
\end{equation} 
where $\Delta \beta=\beta^l-\beta^h$ captures the heterogeneity between the conformity parameters. I use the decomposition $\mathbf{g}_i\boldsymbol\Sigma \mathbf{g}^{\prime}_i=\bar{p}_i^2+\sum_{j}g_{ij}^2p_j(1-p_j)$ which highlights the richness of the model through the quadratic effect of the expected social norm and the role of the social norm's second moment.
Several features of the decision rule in the heterogeneous conformity model are worth highlighting. 

First, assume that $\beta^h=\beta^l>0$, such that there is no heterogeneity in the preference for conformity. In that case, the marginal expected utility only depends on the player's characteristics, $\alpha_i$, and the expected social norm, $\bar p_i$. If player $i$ expects that a majority of her friends play the high action, then conforming to the norm and selecting $y_i=1$ would increase her marginal expected utility by $\beta^h(\bar{p}_i-\frac{1}{2})>0$, a pattern that resembles majority games \citep{jackson2015} and coincides with the homogeneous conformity model of \cite{brock_discrete_2001}.\footnote{In \cite{brock_discrete_2001} and \cite{lee_binary_2014}, the majority threshold is zero instead of $\frac{1}{2}$ and does not appear explicitly, since they use an alternative coding scheme, $y_i=\{-1,1\}$. I study the heterogeneous conformity model resulting from this coding scheme in Online Appendix B.3.} Conversely, if player $i$ expects that a majority of her friends select the low action, then $\bar{p}_i<\frac{1}{2}$, and playing $y_i=1$ would decrease her marginal expected utility by $\beta^h(\bar{p}_i-\frac{1}{2})<0$. If $\beta^h=\beta^l$ are negative, players are anticonformist and prefer to diverge from the norm. As such, selecting $y_i=1$ increases player $i$'s marginal expected utility only if she expects a minority of friends to play the high action.

Second, consider the case in which the tastes for conformity are heterogeneous, so that the marginal expected utility of choosing the high action also depends on the quadratic expected social norm, $\bar p_i^2$, as well as on the variance of the social norm, $\sum_{j\neq i}g_{ij}^2p_j(1-p_j)$. Introducing heterogeneity in conformity preferences thus yields a richer decision rule, in which the marginal expected utility depends nonlinearly on the average behavior of peers and also on the dispersion of peers’ actions. 
For instance, if $\beta^l> \beta^h\geq0$, then $\Delta \beta> 0$, and the marginal effect of the expected norm on the expected utility of choosing the high action is positive since $\frac{\partial \mathbb{E}[\Delta _iU_i]}{\partial \bar p_i}=\beta^h +  \Delta \beta \bar{p}_i$. Moreover, it increases with $\bar{p}_i$, as $\frac{\partial^2 \mathbb{E}[\Delta_iU_i]}{\partial \bar{p}_i^2} =\Delta \beta$. This is intuitive: when the taste for conformity is stronger when choosing the low action, an increase in $\bar{p}_i$ raises the distance to the norm when selecting the low action, thereby strengthening the incentive to choose the high action. 
Conversely, if $\beta^h > \beta^l \geq 0$, then $\Delta \beta < 0$, so the marginal effect of the expected social norm decreases with $\bar{p}_i$, although it remains positive even when $\bar{p}_i=1$.
The sign of $\Delta \beta$ therefore determines whether the marginal expected utility of choosing the high action is a concave or convex function of the expected norm. Note that the nonlinear effect of the expected social norm arises from the combination of heterogeneity and a quadratic social distance function. A linear (albeit heterogeneous) social distance yields a marginal expected utility that is piecewise linear in the expected norm, whereas a homogeneous (but quadratic) specification implies a marginal effect that is linear in $\bar{p}_i$.\footnote{If one considers a linear social distance function with heterogeneous conformity preferences, such as $S^{lin}(y_i,\mathbf{y}_{-i})=y_i\beta^h\left(\bar{y}_i-y_i\right)+ (1-y_i)\beta^l\left(y_i-\bar{y}_i\right)$, the marginal expected utility would be $\mathbb{E}[\Delta _iU_i]=\alpha_i+\left[\beta^h\left(\bar{p}_i-1\right)+\beta^l\bar{p}_i\right]-\eta_i$, which is piecewise linear in $\bar{p}_i$. However, action-specific heterogeneity can still be identified if there are isolated players in the network (see Online Appendix B.2). Assuming a homogeneous quadratic distance function yields $\mathbb{E}[\Delta _iU_i]=\alpha_i+\left\{\beta\left(\bar{p}_i-\frac{1}{2}\right)\right\}-\eta_i$, which is linear in $\bar{p}_i$ (see Online Appendix A.1).}  \\

\noindent \textbf{Best-Response Function}. Let $\Gamma_i(\mathbf p):\{0,1\}^{n} \rightarrow \{0,1\}$ be the best response function, which maps player $i$'s rational expectations about other players' actions, $\mathbf p_{-i}$, into the conditional choice probability, such that $\Gamma_i(\mathbf p)=\Pr(y_i=1\mid \boldsymbol \alpha,\mathbf G)$. Note that under Assumption \ref{H1}, the conditional choice probability $\Pr(y_i=1\mid \boldsymbol \alpha,\mathbf G)$ can be derived from Equation \eqref{deltaU} as:
\begin{equation}\label{Pr}
\Pr(y_i=1\mid \boldsymbol \alpha,\mathbf G)=F_\eta\left(\alpha_i+\beta^h\left(\bar{p}_i-\frac{1}{2}\right)+\frac{1}{2}\Delta \beta \mathbf{g}_i\boldsymbol\Sigma  \mathbf{g}^{\prime}_i\right)
\end{equation}

\section{Equilibrium Properties} \label{section2}
\subsection{Existence of a Bayesian Nash Equilibrium}
 A rational expectations profile $\mathbf{p}$ is \textit{consistent} with respect to the distribution of players' types if $p_i=\Gamma_i(\mathbf p)$, where $p_i=\mathbb{E}[y_i\mid \boldsymbol \alpha,\mathbf G]$ is the expectation of player $i$'s action, as defined in Assumption \ref{H2}.   

The profile $\mathbf p^*$ is a BNE if $p^*_i = \Gamma_i(\mathbf p^*)$ for all $i \in \mathcal{N}$. Indeed, letting $\Gamma=\{\Gamma_i\}_{i \in \mathcal{N}}$, it follows from Brouwer fixed point theorem that $\mathbf p^*$ is a fixed point of $\Gamma$ such that $\Gamma(\mathbf p^*)=\mathbf p^*$. The strategy profile $\mathbf p^*$ at a BNE can be written as:
\begin{equation} \label{eq7}
 \mathbf p^*=F_\eta\left(\boldsymbol\alpha+\beta^h\left(\bar{\mathbf{p}}^*-\frac{1}{2}\mathbf{1}_n\right)+\frac{1}{2}\Delta \beta \operatorname{diag}(\mathbf G\boldsymbol\Sigma ^* \mathbf G^{\prime})\right)
\end{equation}
\noindent where $\boldsymbol\Sigma ^*=\left(\mathbf p^*\mathbf p^{*\prime}+\operatorname{diag}(\mathbf p^*\circ(\mathbf 1_n-\mathbf p^*))\right)$. 

\subsection{Uniqueness of the Bayesian Nash Equilibrium}
To ensure the uniqueness of the equilibrium, it suffices to show that the mapping defined by Equation \eqref{eq7} is a contraction. The following condition therefore bounds the strength of the conformity parameters.\footnote{A simpler but more restrictive sufficient condition for uniqueness can be derived from Assumption \ref{H3}: $\max\{\lvert \beta^h \rvert,\lvert \beta^l \rvert \}<\frac{1}{4\max_u f_{ \eta}(u)}$.}

\begin{assumption}\label{H3}
The conformity parameters are bounded: $ \lvert \beta^h\rvert+\frac{3}{2} \lvert\beta^l-\beta^h\rvert < \frac{1}{\max_u f_{ \eta}(u)}$.
\end{assumption}
 If $f_\eta$ is logistic or normal, Assumption \ref{H3} implies $\lvert\beta^h\rvert+\frac{3}{2} \lvert\beta^l-\beta^h\rvert < 4$ or $\lvert\beta^h\rvert+\frac{3}{2} \lvert\beta^l-\beta^h\rvert < \sqrt{2\pi}$, respectively. The term $\frac{3}{2} \lvert\beta^l - \beta^h\rvert$ in the uniqueness condition shows that heterogeneity in tastes for conformity imposes additional restrictions on the strength of conformity effects. In particular, a one-unit increase in the difference between the two conformity parameters threatens the uniqueness of the equilibrium more than a one-unit increase in $\beta^h$. This is because $\Delta \beta$ affects both the expected social norm, $\bar p_i$, and the variance of peer behavior, $\sum_{j \neq i} g_{ij}^2 p_j(1-p_j)$, both of which are non-negative. Hence, when $\lvert\beta^l - \beta^h\rvert$ is large, multiple equilibria may arise, even if the magnitude of $\lvert \beta^h \rvert$ alone is not sufficient to generate them.
 
 If the preference for conformity is homogeneous, i.e., $\beta^l=\beta^h=\beta$, the sufficient condition for the uniqueness of the equilibrium simplifies to $\lvert \beta \rvert< \frac{1}{\max_u f_{ \eta}(u)}$, which is the same condition as in \cite{lee_binary_2014}. 
  Note that a similar assumption, called Moderate Social Influence, is used to guarantee uniqueness in games with continuous action spaces \citep{horst2006,glaeser2003}. 
 With binary action spaces, incomplete information generally provides straightforward conditions for uniqueness \citep[see, e.g.,][]{lin2024,xu2018,liu2019}. Conversely, binary games with complete information generally have multiple equilibria \citep{depaula2013,ciliberto2009}. For such models, \cite{li2016} and \cite{leung2020} propose promising partial identification approaches using subnetworks and strategic neighborhoods, respectively. Other papers \citep{krauth2006,soetevent2007,bajari2010identification} specify or estimate a selection rule to determine which equilibrium is played to complete the model \citep{tamer2003}. Another option is to assume a dynamic game with myopic players, as in \cite{nakajima2007} and \cite{Badev2021}. I leave the extension of the heterogeneous conformity model to complete information settings for future research. 

\begin{proposition}\label{P1}
Under Assumptions \ref{H1} to \ref{H3}, the network game of incomplete information with utility \eqref{U} has a unique Bayesian Nash equilibrium, given by $\mathbf{p}^*$ in Equation (\ref{eq7}). 
\end{proposition}
\noindent The proof of Proposition \ref{P1} is provided in Appendix \ref{AppendixB} and is based on the contraction property of the mapping $\Gamma(\mathbf p)$. $\Gamma(\mathbf p^*)$ being a contraction mapping is a sufficient condition for the uniqueness of the fixed point $\mathbf p^*=\Gamma(\mathbf p^*)$. As the consistency of the rational expectation belief system implies that for any $\mathbf p$, $\mathbf p=\Gamma(\mathbf p)$ only if $\mathbf p$ is an equilibrium, the unique fixed point $\mathbf p^*$ is the unique equilibrium of the game. Note that Assumption \ref{H3} is a sufficient but not necessary condition for uniqueness (see Section 5.5 in \cite{bhattacharya2024} for a related discussion). 

\subsection{Comparative Statics}
This subsection examines the comparative statics of the equilibrium profile, with a focus on the role of action-specific heterogeneity in conformity preferences. Proofs are omitted because they follow directly directly from total differentiation of the equilibrium profile $\mathbf p^*$ with respect to $\beta^l$, $\beta^h$ and $\bar{\mathbf{p}}^*$. %Note that identification of the parameters in nonlinear peer effects models requires that at least one variable in $\alpha$, say $X_k$, has a large support and a non-zero associated coefficient, $\gamma_k$, so that I have, without loss of generality, $\text{supp}\left(\alpha\right)=(-\infty,\infty)$ \citep{brock2007}.  %I define the global social norm among all players as $\overline{p^*}$.

\begin{corollary} \label{compa}
    \begin{enumerate}[(i)]   
    \item $\operatorname{sign}\left(\mathbf p^*_{(1)} - \mathbf p^*_{(0)}\right)=
\operatorname{sign}\left(\beta^l_{(1)} - \beta^l_{(0)}\right)\mathbf 1_n$
    \item $\operatorname{sign}\left(\mathbf p^*_{(1)} - \mathbf p^*_{(0)}\right)=
\operatorname{sign}\left(\beta^h_{(0)} - \beta^h_{(1)}\right)\mathbf 1_n$
\item $\operatorname{sign}\left(\mathbf p^*_{(1)} - \mathbf p^*_{(0)}\right)
=
\operatorname{sign}\left(\bar{\mathbf p}^*_{(1)} - \bar{\mathbf p}^*_{(0)}\right)
\cdot
\operatorname{sign}\left(\beta^h(\mathbf{1}_n -\bar{\mathbf p})+ 
\beta^l\bar{\mathbf p} \right)$
    \end{enumerate}
\end{corollary}

Corollary \ref{compa}$(i)$ implies that an increase in the conformity taste for conformity when playing the low action always leads to a \textit{higher} equilibrium. Indeed, agents choosing the low action are always weakly \textit{below} the expected norm, and when $\beta^l$ increases, they face a larger penalty from deviating. This may induce them to switch to the high action, thereby yielding a higher equilibrium. Hence, if a social planner wishes to attain a \textit{higher} (\textit{lower}) equilibrium, increasing (decreasing) $\beta^l$ through social marketing campaigns would be an effective policy.

Corollary \ref{compa}$(ii)$ delivers an opposite insight for $\beta^h$. Since agents who play the high action are always weakly \textit{above} the expected norm, increasing (decreasing) $\beta^h$ leads to a larger penalty and thus a \textit{lower} (\textit{higher}) equilibrium. 

Corollary \ref{compa}$(iii)$ highlights the non-monotonic effect of the local norm on players' expected utility and best response strategies. Specifically, the impact of a shift of the expected norm on the equilibrium strategy profile can be positive for some players and negative for others, depending on the idiosyncratic value of the local norm. Conversely, in the homogeneous conformity model, the effect of the expected norm depends only on $\beta$.
 Corollary \ref{compa}$(iii)$ also shows that if $\beta^h$ and $\beta^l$ are both positive (negative), shifting the expected social norm toward the high action, e.g., via a social norm nudge, enables the social planner to reach a \textit{higher} (\textit{lower}) equilibrium, independently of the initial value of the norm. This result is particularly useful as one may expect that the action-specific tastes for conformity are generally of the same sign. However, if the taste for conformity when choosing the high action is negative while the other parameter is positive, a nudge that shifts the norm toward the high action, thereby decreasing the term $\frac{\mathbf{1}_n - \bar{\mathbf p} }{\bar{\mathbf p}}\beta^l$, may lead to a \textit{lower} equilibrium. 

\section{Econometric Model}\label{section3}
This section introduces the econometric model and discusses the identification of the parameters. I also propose an estimation strategy using the Nested-Pseudo Likelihood (NPL) estimator by \cite{aguirregabiria2007}. For consistency with the empirical analysis, which uses network data collected from many independent schools, I assume that the data are drawn from \textit{many networks}.\footnote{To keep the notation as simple as possible, I use the network index $m=1,\dots,M$ only when necessary.} Let $M$ be the number of networks, and $n_m$ be the number of players in network $m$. Accordingly, the asymptotic analysis of the estimator assumes that $M$ goes to infinity, while $n_m$ is bounded.\footnote{In the econometric literature on network games with binary outcomes, an alternative asymptotic framework considers one large network, where the number of players in the network tends to infinity \citep[see, e.g.,][]{lin2017,xu2018,leung2015,menzel2016}.} The original asymptotic framework of \cite{aguirregabiria2007}, where the number of markets grows while the number of players is fixed, parallels the setting here.  
\subsection{Econometric Model}
Let $d_\mathbf{X}$ be the number of columns in the matrix $\mathbf{X}$. I use a standard empirical specification for $\alpha_i$, $\alpha_i=\gamma_0 + \mathbf x_i^{\prime}\boldsymbol\gamma_1 +\bar{\mathbf{x}}_i^{\prime}\boldsymbol\gamma_2$, where $\mathbf x_i\in \mathbb{R}^{d_\mathbf{X}}$ are $i$'s characteristics and $\bar{\mathbf{x}}_i=\sum_{j\neq i}g_{ij}\mathbf x_j \in\mathbb{R}^{d_\mathbf{X}}$ are the average characteristics of $i$'s friends and capture exogenous peer effects. The vector $\mathbf{z}_i=\left(1, \mathbf{x}_i^{\prime}, \bar{\mathbf{x}}_i^{\prime}\right)^{\prime}$ and the matrix $\mathbf{Z}=[\mathbf{z}_1\dots\mathbf{z}_n]'$ thus correspond to the model's control variables and $\boldsymbol\gamma = (\gamma_0, \boldsymbol\gamma_1^{\prime}, \boldsymbol\gamma_2^{\prime})^{\prime}$ are the associated parameters. Observe that network fixed effects are not included in $\alpha_i$ to prevent potential incidental parameters problems when $M$ is large.\footnote{In the empirical application, $n_m$ is large enough so that the incidental parameters problem does not arise \citep{lee_binary_2014} and I thus include in $\alpha_i$ a $M\times 1$ vector of fixed effect to control for correlated effects at the school level.} In addition, let $\boldsymbol \theta=(\boldsymbol \gamma^{\prime},\beta^h, \Delta \beta)^{\prime}$ and the matrix  $\mathbf{K}=[\mathbf{k}_1\dots\mathbf{k}_n]'$, where $\mathbf k_i=\left(\mathbf z_i^{\prime}, \bar{p}_i-\frac{1}{2}, \frac{1}{2}\mathbf g_i\boldsymbol\Sigma  \mathbf g_i^{\prime}\right)^{\prime}$, gather the model's parameters and variables, respectively.
\subsection{Identification}
The identification strategy concerns the parameters $\boldsymbol \theta$ and the distribution function of players' type, $F_\eta$. Because the estimation strategy is parametric in nature, I assume that $F_\eta$ is known.\footnote{I refer the interested reader to \cite{houndetoungan_count_2020} who adapts the identification arguments of \cite{manski1988} to network models with discrete outcomes. Identification of $F_\eta$ holds if there exists a large support variable in $\mathbf x_i$ and if the variance of $\eta_i$ is set to one.} 
Notice that since $\boldsymbol{\alpha}_i=\mathbf{z}_i^{\prime}\boldsymbol \gamma$, $\mathbf{p}_m=\Pr\left(\mathbf{y}_m\mid \mathbf{Z}_m,\mathbf{G}_m;\boldsymbol\theta\right)$.
 Identification fails if two different parameters $\boldsymbol \theta$ and $\tilde{\boldsymbol \theta}$ yield the same equilibrium $\mathbf p_m$ in network $m$, i.e., the two parameters are observationally equivalent (see Definition \ref{def}). Defining observational equivalence at the network level is standard in peer effect models \citep[see, e.g.,][]{guerra2022,brock2007,davezies2009,lee_identification_2007,Bramoulle2009,houndetoungan_count_2020}. Indeed, because $\Pr\left(\mathbf{y}_m\mid \mathbf{Z}_m,\mathbf{G}_m;\boldsymbol\theta\right)$ varies with $\mathbf{G}_m$, observing many networks with different interaction structures provides variation in $\mathbf{G}_m$ across networks that complements the standard variation in $\mathbf{z}_i$ across players and contributes to identification \citep{rothenberg1971}.
 \begin{definition}\label{def}
 Two parameters $\boldsymbol\theta$ and $\boldsymbol{\tilde{\theta}}$ are observationally equivalent conditional on $\mathbf{Z}_m$ and $\mathbf{G}_m$ if $\Pr\left(\mathbf{y}_m\mid \mathbf{Z}_m,\mathbf{G}_m;\boldsymbol\theta\right)=\Pr\left(\mathbf{y}_m\mid \mathbf{Z}_m,\mathbf{G}_m;\boldsymbol{\tilde{\theta}}\right)$ for any network $m$.
 \end{definition}
 The condition $\Pr\left(\mathbf{y}_m\mid \mathbf{Z}_m,\mathbf{G}_m;\boldsymbol\theta\right)=\Pr\left(\mathbf{y}_m\mid \mathbf{Z}_m,\mathbf{G}_m;\boldsymbol{\tilde{\theta}}\right)$ implies that the two distributions yield the same conditional expected outcome $\mathbf{p}_m=\mathbb{E}\left[\mathbf y_m\mid  \mathbf Z_m, \mathbf G_m \right]$ for all $m$. Since $\mathbf{p}_m=\Gamma(\mathbf{K}_m\boldsymbol{\theta})$ and the best-response mapping $\Gamma$ is strictly monotone, observational equivalence requires $\mathbf{K}_m(\boldsymbol{\theta}-\boldsymbol{\tilde{\theta}})=0$ for all $m$, which implies $\boldsymbol{\tilde{\theta}}=\boldsymbol\theta$ whenever the matrix $\mathbf{K}_m$ is full rank. Consequently, $\boldsymbol{\tilde{\theta}}\neq \boldsymbol \theta$ cannot be observationally equivalent to $\boldsymbol \theta$ if the matrix $\mathbb{E}[\mathbf{K}'_m \mathbf{K}_m]$ is full rank. The following assumption formalizes the conditions necessary for identification. 
 
\begin{assumption}\label{H4}
\begin{enumerate}[$(i)$]
\item  $\mathbf{y}_m$, $\mathbf{Z}_m$ and $\mathbf{G}_m$ are observed for many networks $m=1,\dots,M$.
    \item The matrix $\operatorname{plim}_{M \to \infty}\frac{1}{M}\sum_{m=1}^M\sum_{i=1}^{n_m} \mathbf{k}_i \mathbf{k}_i^{\prime}$ is full rank.
\end{enumerate}
 
\end{assumption}
Assumption \ref{H4}$(i)$ defines what is observed by the econometrician. Assumption \ref{H4}$(ii)$ involves the probability limit of the sample analogue of $\mathbb{E}\left[\sum_{i=1}^{n_m} \mathbf{k}_i \mathbf{k}_i^{\prime}\right]$, to which it converges as the number of networks goes to infinity by the law of large numbers. Note that this type of assumption is standard in peer effects models \citep[see, e.g.,][]{graham2005,lee_identification_2007,liu2017,houndetoungan_count_2020} and fails if the model's explanatory variables are linearly dependent.\footnote{The presence of the constant $\frac{1}{2}$ in the term $\beta^h\left(\bar{p}_i-\frac{1}{2}\right)$ does not threaten the identification of $\gamma_0$, even if there are no isolated players, in which case the intercept is $\gamma_0-\frac{\beta^h}{2}$. Indeed, $\beta^h$ is identified independently by $\bar{p}_i$ if Assumption \ref{H4} holds.} Moreover, having certain networks for which $\sum_{i=1}^{n_m} \mathbf{k}_i \mathbf{k}_i^{\prime}$ is not full rank does not violate Assumption \ref{H4}$(ii)$, because observing many networks ensures that the average independent variation across networks is sufficient for identification.
 %Especially, the second-to-last element of $\mathbf{k}_i$, $\bar{p}_i-\frac{1}{2}$, varies across players in a given network if there exist two players $i$ and $j$ who face a different social norm, which is likely to hold except if all networks are complete or all players have the same characteristics.
 %Similarly, $\mathbf g_i\boldsymbol\Sigma  \mathbf g_i^{\prime}$ varies across players in a given network if there exist two players $k$ and $\ell$ for which $\mathbf{g}_k \boldsymbol \Sigma  \mathbf{g}_k^{\prime} - \mathbf{g}_\ell \boldsymbol \Sigma  \mathbf{g}_\ell^{\prime}\neq 0 \iff   (\bar{p}_k^2 - \bar{p}_\ell^2) + \sum_{r=1}^n(g_{kr} - g_{\ell r})p_j(1-p_j)\neq 0$. This fails only if the two terms exactly cancel each other, which is unlikely to happen for all pairs of players in every network, or if $\bar{p}_k=\bar{p}_\ell$, which is unlikely to hold for every pair of players as previously discussed. 
 Especially, since $\bar{p}_i-\frac{1}{2}$ and $\frac{1}{2}\mathbf g_i\boldsymbol\Sigma  \mathbf g_i^{\prime}$ both depend on $\mathbf{p}_m$ and $\mathbf{G}_m$, there must be a strictly positive proportion of networks in which at least one player has more than one friend. Indeed, if all players have at most one friend, then any nonisolated player $i$ has only one friend $k$, such that $\bar{p}_i=p_k$. Hence, $\mathbf{g}_i \boldsymbol \Sigma \mathbf{g}_i^{\prime}=p_k^2+p_k(1-p_k)=\bar{p}_i$, which implies that $\bar{p}_i-\frac{1}{2}$ and $\frac{1}{2}\mathbf g_i\boldsymbol\Sigma  \mathbf g_i^{\prime}$ would be linearly dependent.

\begin{proposition}\label{P2}
Under Assumption \ref{H4}, the parameters $\boldsymbol \theta$ of the heterogeneous conformity model given by Equation (\ref{eq7}) are identified. 
\end{proposition}
\noindent The proof of Proposition \ref{P2} is standard in network games and is provided in Appendix \ref{AppendixC}. \\ 

A limitation of Proposition \ref{P2} is that the rank condition in Assumption \ref{H4}$(ii)$ is imposed on variables that depend on $\mathbf p_m$. This equilibrium quantity is not observed in the data.
In his conformity model for count outcomes, \cite{houndetoungan_count_2020} proposes a solution built on the existence of intransitive triads, following the seminal identification strategy of \cite{Bramoulle2009}. An intransitive triad is a subnetwork configuration in which a given agent $i$ has friends of friends who are not her direct friends, that is, $i$ is friend with $j$ who is friend with $k$, but $i$ is not friend with $k$. Because this paper's model includes action-specific conformity parameters, identification requires the existence of \textit{double} intransitive triads, in which players $i$ and $k$ are not friends but share two common friends (Figure \ref{fig:double_triad}).

\begin{figure}[H]
    \centering
    \begin{tikzpicture}[>=stealth]
    \tikzset{
        nodeh/.style={circle, draw, fill=gray!20, thick, inner sep=1pt}
    }

    \node[nodeh] (i) at (0,2)   {$i$};
    \node[nodeh] (j) at (-1.5,0) {$j$};
    \node[nodeh] (l) at (1.5,0)  {$\ell$};
    \node[nodeh] (k) at (0,-2)  {$k$};

    % First path
    \draw[->, thick, blue] (i) -- (j);
    \draw[->, thick, blue] (j) -- (k);

    % Second path
    \draw[->, thick, red] (i) -- (l);
    \draw[->, thick, red] (l) -- (k);

    \end{tikzpicture}

    \vspace{0.5em}
    \begin{minipage}{0.85\textwidth}
        \small\textit{Note:} This figure depicts the minimal link structure required in Condition $(ii)$ of Proposition \ref{PropoAri}. The only forbidden link is $g_{ik}=0$. All other links not depicted in the figure, namely $g_{ji}$, $g_{kj}$, $g_{k\ell}$, $g_{\ell i}$, $g_{j\ell}$, $g_{\ell j}$, and $g_{ki}$, are left unconstrained and may or may not exist without affecting identification. In undirected networks, all links among $\{i,j,\ell,k\}$ may exist except the link between $i$ and $k$.
    \end{minipage}
      \caption{A \textit{double} intransitive triad: $i$ and $k$ share two common friends, $j$ and $\ell$, but $k$ is not a friend of $i$.}
       \label{fig:double_triad}
\end{figure}

\begin{proposition} \label{PropoAri} Assume that $\beta^h,\beta^l \geq 0$. Assumption \ref{H4}$(ii)$ is satisfied if:
\begin{enumerate}[(i)] 
    \item The matrix $\operatorname{plim}_{M \to \infty}\frac{1}{M}\sum_{m=1}^M\sum_{i=1}^{n_m} \mathbf{z}_i\mathbf{z}_i^{\prime}$ is full rank;
        \item There exists a positive proportion of networks in which some agents $i$ have two friends $j$ and $\ell$ who are both friends with an agent $k$ who is not a friend of $i$;
        \item There is a variable $x_{i,\kappa}$ in $\mathbf{x}_i$ such that $\gamma_{1,\kappa} \gamma_{2,\kappa} \geq 0$ and $\gamma_{2,\kappa} \neq 0$, where $\gamma_{1,\kappa}$ is the coefficient of $\mathbf{x}_{i,\kappa}$ and $\gamma_{2,\kappa}$ is the coefficient of $\bar{\mathbf{x}}_{i,\kappa}$.
\end{enumerate}
\end{proposition}

Condition $(i)$ only imposes that the observed variables in $\mathbf z_i$ are not linearly dependent on average across networks and is thus less strict than Assumption \ref{H4}$(ii)$. Condition $(ii)$ guarantees the existence of \textit{double} intransitive triads across networks. Double intransitive triads are less common than single ones since they require that $i$ has at least two distinct friends, $j$ and $\ell$, that are also friends with $k$, instead of only one. However, given the strong transitivity typically observed in social networks \citep{watts1998,jackson2007}, the number of double intransitive triads is of the same order of magnitude as that of single intransitive triads. In the Add Health data used in the empirical application, 74,783 students generate 213,643 dyads, 852,456 single intransitive triads, and 227,219 double intransitive triads. Condition $(ii)$ is therefore not substantially more restrictive than identification conditions based on the presence of single intransitive triads, as in \cite{Bramoulle2009} and \cite{houndetoungan_count_2020}.
%However, conditional on $d_i \geq 2$ for all relevant agents, double intransitive triads are not necessarily rarer than single ones. The reason is that the scarcity of \textit{any} intransitive configuration in social networks stems primarily from clustering, the tendency for an agent's friends to be mutually linked. Consequently, whenever agents $i$ and $k$ are not connected but share a common neighbor $j$ (a single intransitive triad), transitivity of social ties implies that a second common neighbor $\ell$ is also likely to exist, provided $d_i, d_k \geq 2$. Define the event $A_{j}(i,k) = \{ g_{ij}>0 \wedge g_{kj}>0 \}$ that $j$ is a common friend of $i$ and $k$. Clustering implies that link formation is positively correlated across agents: for any $\ell \neq j$,
%\begin{equation*}
 %   \Pr\left( A_{\ell}(i,k)  \mid A_{j}(i,k),\, g_{ik} = 0 \right)   >    \Pr \left( A_{\ell}(i,k) \mid g_{ik} = 0 \right)>0,
%\end{equation*}
%since the event $A_{j}(i,k)$ is evidence that $i$ and $k$ belong to overlapping social circles, raising the likelihood that any other agent $\ell$ is also a friend of both. The last inequality holds because $\Pr \left( A_{\ell}(i,k) \mid g_{ik} = 0 \right)$ is the probability that a single intransitive triad exists, where $k$ is a friend of friend of $i$ through $\ell$. The probability that a double intransitive triad, in which $i$ has two friends $j$ and $\ell$ who share a common friend $k$ who is not herself a friend of $i$, is strictly positive if an intransitive triad between $i$, $j$ and $k$ exists. 
Condition $(iii)$ guarantees that $p_i$ is correlated with the contextual variable $\bar{\mathbf{x}}_{i,\kappa}$, such that the marginal effect $\frac{d p_i}{d \mathbf{x}_{j,\kappa}}$ is not zero when $\gamma_{2,\kappa} \neq 0$ and $j$ is a friend of $i$. 
The restriction $\gamma_{1,\kappa} \gamma_{2,\kappa} \geq 0$ ensures that the own and contextual effects of $\mathbf{x}_{\kappa}$ operate in the same direction, so that indirect effects through endogenous peer effects reinforce rather than offset the direct effects. This condition holds, for instance, when the effects of a player's age and the average age of her friends on her best response are either both positive or both negative. \\

I present the intuition using a single intransitive triad, where $i$ and $k$ are friends with $j$ but not friends with themselves, while the formal proof, which relies on double intransitive triads, is given in Appendix \ref{AppendixD}. The proof proceeds by contradiction. Assume that Assumption \ref{H4}$(ii)$ is not verified, which implies that the variables in $\mathbf k_i$ are linearly dependent. For player $i$, I can then write: 
\begin{equation}\label{idAri}
   \check{\beta}^h\left(\bar{p}_i-\frac{1}{2}\right)+\frac{1}{2}\widecheck{\Delta\beta} \mathbf g_i\boldsymbol\Sigma  \mathbf g_i^{\prime}=\check{\gamma}_0 + \mathbf{x}_i^{\prime}\check{\boldsymbol\gamma}_1 +\bar{\mathbf{x}}^{\prime}_i\check{\boldsymbol\gamma}_2
\end{equation} 
for some constants $\check{\beta}^h\in \mathbb{R}_{+}$, $\widecheck{\Delta\beta}\in \mathbb{R}$, $\check{\gamma}_0\in \mathbb{R}$, $\check{\boldsymbol{\gamma}}_1\in \mathbb{R}^{d_\mathbf{X}}$ and $\check{\boldsymbol{\gamma}}_2\in \mathbb{R}^{d_\mathbf{X}}$. Since $j$ is the only friend of $i$, Equation \eqref{idAri} simplifies to $ p_j=\left(\check{\beta}^h+\frac{\widecheck{\Delta \beta}}{2}\right)^{-1}\left(\check{\gamma}_0 +\frac{\check{\beta}^h}{2} + \mathbf x_i^{\prime}\check{\boldsymbol\gamma}_1 +\mathbf{x}_j^{\prime}\check{\boldsymbol\gamma}_2\right)$.\footnote{If $j$ is the only friend of $i$, $g_{ij}=1$. Hence $\mathbf g_i\boldsymbol\Sigma  \mathbf g_i^{\prime}=p_j$, $\bar{p}_i=p_j$ and $\bar{\mathbf{x}}_i=\mathbf{x}_j$.}
Thus, $p_j$, the expected action of $j$, is not influenced by $x_{k,\kappa}$ and the marginal effect $\frac{d p_j}{d x_{k,\kappa}}$ is zero. However, since $j$ and $k$ are peers and $\gamma_{2,\kappa} \neq 0$, the marginal effect of $x_{k,\kappa}$ on $p_j$ cannot be zero and there is a contradiction. Under the conditions given in Proposition \ref{PropoAri}, Equation \eqref{idAri} cannot hold for all players when many networks include intransitive triads and $\operatorname{plim}_{M \to \infty}\frac{1}{M}\sum_{m=1}^M\sum_{i=1}^{n_m} \mathbf{z}_i\mathbf{z}_i^{\prime}$ has full rank.

The proof of Proposition \ref{PropoAri} assumes that the heterogeneous conformity parameters are non-negative, which guarantees that the total effect of $x_{k,\kappa}$ on $p_j$ is nonzero (see Lemma \ref{Lemma1} in Appendix \ref{PropoAri}). If $\beta^h$ and/or $\beta^l$ are negative, the direct effects of $x_{k,\kappa}$ via $\gamma_{1,\kappa}$ and $\gamma_{2,\kappa}$ and the indirect effect via endogenous peer effects may have opposite signs. If both types of effects perfectly cancel each other out, the marginal effect of $x_{k,\kappa}$ on $p_j$ would be zero. However, such a perfect offset is highly unlikely to hold for every pair of players across many networks.
    
\subsection{Estimation}
This section presents the strategy for estimating the model parameters $\boldsymbol \theta$, assuming that the researcher observes $\mathbf{y}_m$, $\mathbf Z_m$ and $\mathbf G_m$ for $m=1,\dots,M$. Under Assumption \ref{H2}, players form rational expectations based solely on public information $\{\mathbf Z, \mathbf G\}$. Furthermore, under Assumption \ref{H1}, private payoff shocks are mutually independent. Therefore, conditional on the public information set and the consistent rational expectation profile, each player's action is statistically independent. The conditional log-likelihood function of the observed action profile $\mathbf y$, given $\Gamma(\boldsymbol \theta,\mathbf p^*)=\Pr(\mathbf{y}\mid \mathbf Z,\mathbf G;\boldsymbol \theta)$, is then expressed implicitly as $\mathcal{L}_M(\boldsymbol \theta; \mathbf p^*)$: 
\begin{equation*}\label{LL}
  \mathcal{L}_M(\boldsymbol \theta; \mathbf p^*)
 %=\ln\left(\prod_i^n (p^*_i)^{y_{i}} (1-p^*_i)^{1-y_{i}}\right) \\
 =\frac{1}{M}\sum_{m=1}^M\sum_{i=1}^{n_m} \left\{y_i\times \ln \Gamma_i(\boldsymbol \theta,\mathbf p^*)+(1-y_i)\times \ln(1-\Gamma_i(\boldsymbol \theta,\mathbf p^*))\right\} 
\end{equation*}
\noindent Let $\Theta$ denote the support set of $\boldsymbol \theta$. The MLE is $\hat{\boldsymbol \theta}_{MLE}=\operatorname{argmax}_{\theta \in \Theta} \mathcal{L}_M(\boldsymbol \theta; \mathbf p^*)$ s.t. $\mathbf p^*=\Gamma(\boldsymbol \theta,\mathbf p^*)$, as defined in Equation (\ref{eq7}).\footnote{To estimate standard errors for $\beta^l$ rather than $\Delta \beta$, I use an alternative but equivalent form of Equation (\ref{eq7}): $\Gamma(p^*)=F_{ \eta}\left(\boldsymbol \alpha+ \beta^h\left(\bar{\mathbf{p}}^*-\frac{1}{2}\mathbf{1}_n-\frac{1}{2}\operatorname{diag}\left(\mathbf{G} \boldsymbol\Sigma  \mathbf{G}^{\prime}\right)\right)+\frac{1}{2}\beta^l\operatorname{diag}\left(\mathbf{G} \boldsymbol\Sigma  \mathbf{G}^{\prime}\right)\right)$. For simplicity, I retain the original form of Equation (\ref{eq7}) elsewhere in the paper.} \\ 

However, $\Gamma_i(\boldsymbol \theta,\mathbf p^*)$ is not directly observed in the data and must be obtained as a fixed point solution $p_i^*=\Gamma_i(\boldsymbol \theta,\mathbf p^*)$. To address this issue, I substitute the unobserved equilibrium vector $\mathbf{p}^*$ with an arbitrary vector $\mathbf{p}$ to compute the conditional pseudo log-likelihood $\Tilde{\mathcal{L}}_M(\boldsymbol \theta; \mathbf p)$:
\begin{equation}\label{pseudoLL}
 \Tilde{\mathcal{L}}_M(\boldsymbol \theta; \mathbf p)=\frac{1}{M} \sum_{m=1}^M \sum_{i=1}^{n_m}{\left\{y_i\times \ln\Gamma_i(\boldsymbol \theta,\mathbf p)+(1-y_i)\times \ln(1-\Gamma_i(\boldsymbol \theta,\mathbf p))\right\}} 
\end{equation}
Define $\Tilde{\boldsymbol \theta}_M(\mathbf p)=\operatorname{argmax}_{\boldsymbol \theta \in \Theta} \Tilde{\mathcal{L}}_M(\boldsymbol \theta;\mathbf p)$ and $\Gamma_M(\mathbf p)=\Gamma(\Tilde{\boldsymbol \theta}_M,\mathbf p)$. Two main approaches have emerged in the literature on estimating network games with incomplete information. The first one, used by \cite{lee_binary_2014} and \cite{lin2021}, adapts the Nested Fixed-Point Likelihood method of \cite{rust1987}. For each candidate value of $\boldsymbol \theta$, this sequential approach solves a fixed-point problem to compute $p$ and the associated pseudo-likelihood. However, solving the fixed point at every candidate $\boldsymbol \theta$ is computationally intensive, especially for large $M$. \cite{aguirregabiria2007} propose an alternative, the Nested Pseudo-Likelihood estimator, which updates the fixed point rather than solving it exactly. Define an NPL fixed point as a pair $(\boldsymbol \theta,\mathbf p)$ s.t. $\mathbf p=\Gamma_M(\mathbf p)$. The set of NPL fixed points is given by $\mathcal{A}_M=\{(\boldsymbol \theta,\mathbf p) \in \mathcal{\theta}\times \mathcal{P}: \boldsymbol \theta=\Tilde{\boldsymbol \theta}_M(\mathbf p),\mathbf p=\Gamma_M(\mathbf p) \}$. The NPL estimator is then $(\hat{\boldsymbol \theta}_{NPL},\hat{\mathbf p}_{NPL})=\operatorname{argmax}_{(\boldsymbol \theta,\mathbf p) \in \mathcal{A}_M}\Tilde{\mathcal{L}}_M(\boldsymbol \theta;\mathbf p)$. The sequential estimation starts with an initial guess of $\mathbf p^{(0)}$, then estimates $\hat{\boldsymbol \theta}^{(1)}=\operatorname{argmax}_{\boldsymbol \theta \in \Theta} \Tilde{\mathcal{L}}_M(\boldsymbol \theta; \hat{\mathbf p}^{(0)})$, and updates the fixed point in one step as $\hat{\mathbf p}^{(1)}=\Gamma_M\left(\hat{\boldsymbol \theta}^{(1)},\hat{\mathbf p}^{(0)}\right)$. The estimated $\hat{\mathbf p}^{(1)}$ is then substituted into the pseudo log-likelihood function to obtain updated parameter estimates, $\hat{\boldsymbol \theta}^{(2)}=\operatorname{argmax}_{\boldsymbol \theta \in \Theta} \Tilde{\mathcal{L}}_M(\boldsymbol \theta; \hat{\mathbf p}^{(1)})$. This sequence is repeated until convergence, i.e., when $ \left\| \hat{\mathbf p}^{(k+1)}-\hat{\mathbf p}^{(k)}\right\| <c$ or $\left\|\hat{\boldsymbol \theta}^{(k+1)}-\hat{\boldsymbol \theta}^{(k)}\right\|<c$, where $c$ is a tolerance value sufficiently close to zero. $\boldsymbol \theta^{(k+1)}$ maximizes the pseudo log-likelihood and $\hat{\mathbf p}^{(k+1)}$ is, by construction, a fixed point that satisfies the consistency condition on rational expectations.

 As noted by \citet{aguirregabiria2007}, convergence to a fixed point in the NPL estimator is not theoretically guaranteed. However, \citet{kasahara2012} shows that convergence is guaranteed if the initial starting point lies in the neighborhood of the true parameter and the fixed point mapping is a contraction, the latter holding under Assumption \ref{H3}.\footnote{In the empirical section, I verify the convergence of the NPL estimator by also estimating the Nested Fixed-Point Likelihood estimator.} \\

Under a set of regularity assumptions, the large-sample properties of the NPL estimator, namely, $\sqrt{n}$-consistency and asymptotic normality, are established in \cite{aguirregabiria2007}. Let $\nabla_{k}\Tilde{\mathcal{L}}_M$ be the Jacobian matrix $\nabla_{k} \Tilde{\mathcal{L}}_M(\boldsymbol \theta_0; \mathbf p^*)$, where $\nabla_k$ is the partial derivative with respect to $k$, and $\nabla_k\Gamma$ be the Jacobian matrix $\nabla_k\Gamma(\boldsymbol \theta_0,\mathbf p^*)$. In addition, I define the following matrices: $\Omega_{\boldsymbol\theta \boldsymbol\theta^{\prime}}=\mathbb{E}\left[ \left(\nabla_{\boldsymbol \theta}\Tilde{\mathcal{L}}_M\right) \left(\nabla_{\boldsymbol \theta}\Tilde{\mathcal{L}}_M\right)^{\prime}\right]$, and $\Omega_{\boldsymbol \theta \mathbf p^{\prime}}=\mathbb{E}\left[ \left(\nabla_{\boldsymbol \theta}\Tilde{\mathcal{L}}_M\right) \left(\nabla_{\mathbf p}\Tilde{\mathcal{L}}_M\right)^{\prime}\right]$. The asymptotic distribution of the NPL estimator is given by $\sqrt{M}(\hat{\boldsymbol\theta}_{NPL}- \boldsymbol\theta_0) \overset{d}{\to} \mathcal{N}(0,\mathbb{V}_{NPL})$ where \begin{equation}\label{V}
    \mathbb{V}_{NPL}=[\Omega_{\theta \theta^{\prime}} + \Omega_{\boldsymbol\theta \mathbf p^{\prime}}  ( I - \nabla_{\mathbf p} \Gamma^{^{\prime}})^{-1}  \nabla_{\boldsymbol \theta} \Gamma]^{-1}\Omega_{\boldsymbol \theta \boldsymbol \theta^{\prime}}[\Omega_{\boldsymbol \theta \boldsymbol\theta^{\prime}} + \nabla_{\boldsymbol \theta} \Gamma^{\prime} ( I - \nabla_p \Gamma)^{-1} \Omega_{\boldsymbol \theta \mathbf p^{\prime}}^{\prime}  ]^{-1}
\end{equation} 
\section{Empirical Application}\label{section4}
\subsection{Data}

The empirical application leverages data from the National Longitudinal Survey of Adolescent to Adult Health (Add Health), a widely used source for studying social interactions through detailed friendship networks. The dataset covers 126 U.S. middle and high school students, capturing a broad range of behaviors and socio-demographic characteristics.  The social networks are based on self-reported friendships within the same school and grade, with students allowed to nominate up to five male and five female friends. Importantly, \cite{boucheraristide} show that this censoring of network links does not qualitatively affect the estimate of the peer effect coefficients.  %This results in 532 distinct school-grade networks. 
In addition, participation in the survey was mandatory unless a parent opted out in writing, ensuring high participation rates, which mitigates concerns about measurement error due to missing agents \citep{chandrasekhar2011}. Finally, \cite{hsieh2020} and \cite{Badev2021} show that biases arising from network endogeneity are minimal in the Add Health study. 

The main reason for using the Add Health data in the empirical application is to allow a direct comparison between the heterogeneous peer effects model and the homogeneous model of \cite{lee_binary_2014}. I therefore focus on the same sample, networks, and socio-demographic variables as \cite{lee_binary_2014}, who study smoking behavior. For descriptive statistics of the sample, I refer the reader to their paper.

A student is classified as a non-smoker if they report never smoking or having smoked only once or twice in the past year. I extend this analysis to also examine alcohol consumption, defining a non-drinker as a student who reports never consuming beer, wine, or liquor, or having done so only once or twice in the past year. 23.1\% of students smoke while 55.4\% consume alcohol.
%I analyze two behaviors as I expect to uncover different forms of asymmetries in the costs associated with deviations from the norm. In the case of smoking, I hypothesize that \textit{overdoing} (smoking among non-smokers) incurs heavier social costs than \textit{underdoing} (not smoking among smokers). Conversely, for alcohol consumption, I anticipate that \textit{underdoing} is also costly, as not consuming alcohol among alcohol consumers may be as ostracizing as consuming alcohol among non-consumers. Indeed, smoking is a minority behavior (23.1\% of students smoke in the sample) while alcohol consumption is more common (55.4\%), which shapes the strength and direction of the asymmetries in social costs of deviating, as \textit{overdoing} is more common for smoking than alcohol consumption in this setting.
%The data is drawn from the first wave of the In-School Add Health survey, conducted in 1994-1995, which collected information on the smoking and alcohol consumption of 74,783 students across 127 schools. 

\subsection{Estimation Results}
As a proof of concept, I compare the homogeneous conformity model, used, for instance, in \cite{lee_binary_2014}, with the action-specific heterogeneous model developed in this paper. %Then, I present estimation results that account for the potential misclassification of the binary outcome.
Table \ref{estimation} reports the estimated conformity parameters for the homogeneous and action-specific heterogeneous models, as defined in Equation (\ref{eq7}). To ensure comparability with \cite{lee_binary_2014},\footnote{\cite{lee_binary_2014} encode the binary action as $\mathcal{Y}_i=\{-1,1\}$, while I use $\mathcal{Y}_i=\{0,1\}$. The reported parameters in the first column of Table \ref{estimation} are, by construction, four times larger in my paper but otherwise correspond to Model 6 in \cite{lee_binary_2014}.} I assume that $f_\eta$ is logistic and specify $\alpha_i=m_i \boldsymbol \gamma_0 + \mathbf x_i^\prime \boldsymbol \gamma_1+\bar{\mathbf x}_i\prime \boldsymbol \gamma_2$ where $m_i$ indicates the school of student $i$, i.e., I control for school fixed effects, students' characteristics and exogenous peer effects.
%In addition, I estimate Model 6 in \cite{lee_binary_2014}, which uses a homogeneous model of peer effects. %Similarly, as their marginal effects are computed on a shift from $-1$ to one rather than from zero to one, they are two times larger by definition.}.
\begin{center}
\begin{threeparttable}
\singlespacing
\begin{tabular}{c cc ccc}
\toprule
 & \multicolumn{2}{c}{Homogeneous model} & \multicolumn{3}{c}{Heterogeneous model} \\
\cmidrule(lr){2-3} \cmidrule(lr){4-6}
Conformity parameters & \multicolumn{2}{c}{$\beta$} & $\beta^h$ & $\beta^l$ &$\Delta \beta$\\ 
\cmidrule(lr){1-1} \cmidrule(lr){2-3} \cmidrule(lr){4-6}
\textit{Smoking} & \multicolumn{2}{c}{$2.392^{***}$} & $1.077^{***}$ & $3.980^{***}$ &$2.903^{***}$\\ 
 & \multicolumn{2}{c}{(0.155)} & (0.265) & (0.237) &(0.436)\\ 
\textit{Alcohol Consumption} & \multicolumn{2}{c}{$1.788^{***}$} & $1.772^{***}$ & $1.806^{***}$ &$0.035$\\ 
 & \multicolumn{2}{c}{(0.156)} & (0.198) & (0.208) &(0.284)\\ 
 \cmidrule(lr){1-1} \cmidrule(lr){2-3} \cmidrule(lr){4-6}
Own Effects & \multicolumn{2}{c}{X} & \multicolumn{3}{c}{X} \\ 
Exogenous Peer Effects & \multicolumn{2}{c}{X} & \multicolumn{3}{c}{X} \\ 
School Fixed Effects & \multicolumn{2}{c}{X} & \multicolumn{3}{c}{X} \\ 
\cmidrule(lr){1-1} \cmidrule(lr){2-3} \cmidrule(lr){4-6}
Observations & && & \\ 
\textit{Smoking} & \multicolumn{2}{c}{74,783} & \multicolumn{3}{c}{74,783} \\ 
\textit{Alcohol Consumption} & \multicolumn{2}{c}{74,377} & \multicolumn{3}{c}{74,377} \\ 
\cmidrule(lr){1-1} \cmidrule(lr){2-3} \cmidrule(lr){4-6}
Log-Likelihood && & & \\ 
\textit{Smoking} & \multicolumn{2}{c}{-36,918.78} & \multicolumn{3}{c}{-36,878.35} \\ 
\textit{Alcohol Consumption} & \multicolumn{2}{c}{-46,992.86} & \multicolumn{3}{c}{-46,992.85} \\ 
\bottomrule
\end{tabular} 
\begin{tablenotes} 
\item \footnotesize \textit{Note}: $^{*} p-value < 0.05$, $^{**} p-value < 0.01$, $^{***} p-value< 0.001$. The number of observations differs between the two behaviors because students may choose not to answer certain survey questions.
\end{tablenotes} 
\caption{Structural estimation of the conformity parameters}
\label{estimation}
\end{threeparttable}
\end{center}
 For both behaviors under scrutiny, the heterogeneous model reaches the highest likelihood, although likelihood ratio (LR) tests between the homogeneous and heterogeneous models indicate a significantly higher likelihood for smoking behavior only ($p-value<0.001$). Since the preference for conformity is homogeneous for alcohol consumption, the specification test proposed in Online Appendix A.3 can only be used for smoking. The null hypothesis that the pure conformity model is consistent with the data cannot be rejected at a significance level below 0.2\% ($p-value=0.002$), while the null hypothesis that the pure spillover is consistent with the data is strongly rejected ($p-value<10^{-16}$).\footnote{Note that since the sample size if large (74,783), I favor a strict significance level to reduce the risk of Type I error. At the 5\% or 1\% significance level, the specification test indicates that the pure conformity model is not consistent with the data.}  %Note that the sufficient condition for equilibrium uniqueness stated in Assumption \ref{H3} is not satisfied in the case of smoking, given the estimated parameters. However, the bound on the strength of the social cost parameters imposed by Assumption \ref{H3} represents a worst-case scenario,  with the advantage of depending solely on those parameters. Empirically, I can verify that $\Gamma(\boldsymbol \theta,\mathbf p^*)$ is contracting, given the observed network and individual characteristics, even when Assumption \ref{H3} is violated.

\begin{figure}[H]
\begin{centering}
\subfloat{\label{a}\includegraphics[width=.50\linewidth]{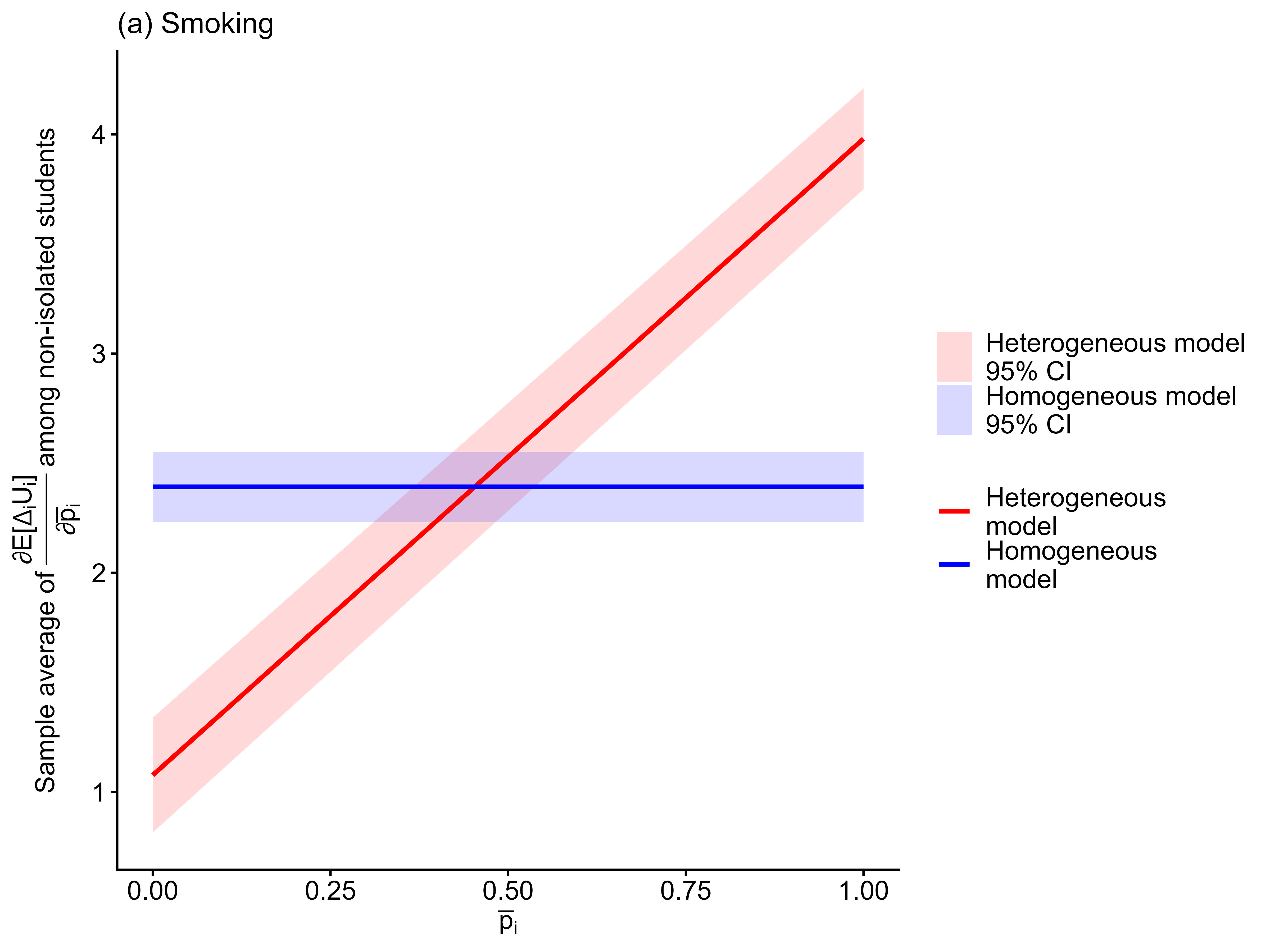}}\hfill
\subfloat{\label{b}\includegraphics[width=.50\linewidth]{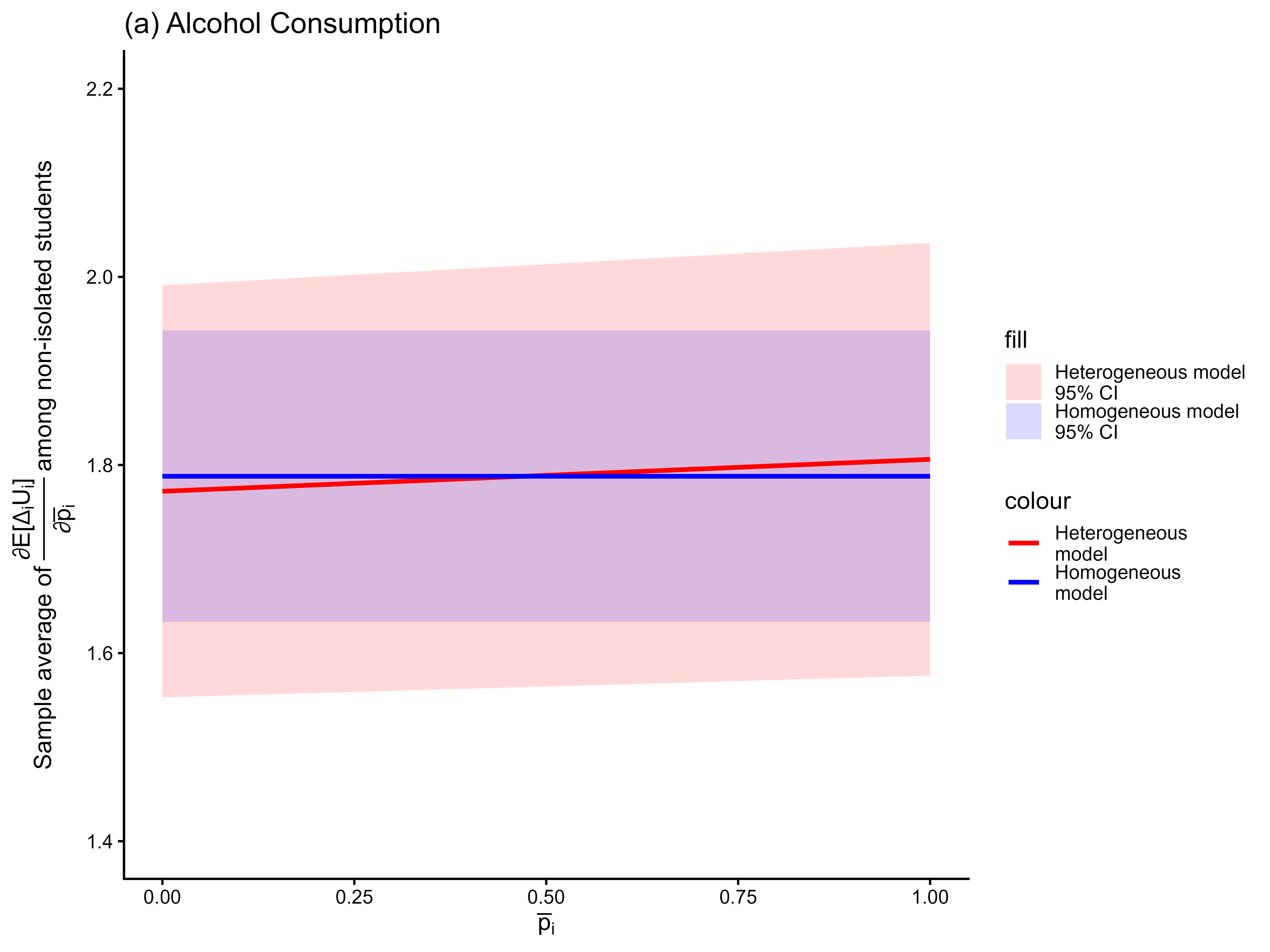}}
\caption{Effect of the expected norm on the marginal expected utility $\mathbb{E}[\Delta _iU_i]$} \label{fig3}
\end{centering}
\medskip
\footnotesize The red line is the sample average marginal effect of the expected norm $\bar{p}$ on $\Delta\mathbb{E}[U]$ among non-isolated students in the heterogeneous model. Since $\Delta \beta$ and $\beta^h$ are positive for both behaviors, the marginal effect increases with the norm in the heterogeneous specification. The blue line corresponds to the homogeneous model, where the marginal effect is constant and equal to $\beta$.
\end{figure}

In the case of smoking, comparing the estimated coefficients in both models reveals a pronounced action-specific heterogeneity in the preference for conformity, not accounted for by the homogeneous conformity parameter $\beta$. Specifically, in the homogeneous specification for smoking behavior, $\widehat{\beta}$ is estimated at $2.392$. In contrast, the heterogeneous model yields a lower estimate for taste for conformity when choosing the high action, $\widehat{\beta}^h=1.077$, and a larger estimate for the taste for conformity when choosing the low action, $\widehat{\beta}^l=3.980$. %Following Proposition \ref{P0}, $\hat{\lambda}=0.45$ and $\beta$ is a weighted average of $\beta^h$ and $\beta^l$.
The heterogeneous model implies that the preference for conformity is stronger among non-smokers than among smokers. This result is intuitive: choosing not to smoke despite having at least one friend who smokes imposes a double penalty, both from deviating from the norm and exposure to passive smoking. Given the addictive nature of smoking, it is not surprising that smokers are relatively unaffected by social norms. Note that $\widehat{\Delta \beta}$ is positive and significant, indicating a positive and convex effect of the expected social norm on the marginal expected utility of smoking. Panel (a) of Figure \ref{fig3} illustrates how the marginal effect of the expected norm on the expected utility of smoking rather than not smoking, $\frac{\partial \mathbb{E}[\Delta _iU_i]}{\partial \bar{p}_i}$, increases as the expected norm approaches one in the heterogeneous model. As smoking becomes more prevalent among one’s friends, abstaining becomes increasingly costly, since choosing an action \textit{below} the norm is more penalized than choosing an action \textit{above}. Conversely, in the homogeneous model, the marginal effect of the norm is constant: a 10pp shift of the expected norm from 10 to 20\% has the same effect as a shift from 80 to 90\%.
In contrast to smoking, the estimated heterogeneous tastes for conformity for drinking alcohol are not significantly different, suggesting that individuals have a homogeneous preference for conformity across actions. In this case, the specification with a homogeneous social distance function provides a satisfactory approximation (see Panel (b) of Figure \ref{fig3} for an illustration). %Here, $\hat{\lambda}=0.46$ and $\beta$ is also a weighted average of $\beta^h$ and $\beta^l$.

\begin{comment}
\begin{center}
\begin{threeparttable}
\singlespacing
\begin{tabular}{cccc}
\toprule
%& {Reduced-form model} & \multicolumn{2}{c}{Structural model} \\ 
% \cmidrule(lr){2-2} \cmidrule(lr){3-4} 
% Average Marginal Effect (AME) & {$\beta$} & {$\beta^h$} & {$\beta^l$} \\ \hline
%\textit{Smoking} & $0.272^{***}$ & $0.393^{***}$ & $0.008$ \\
%& (0.019) & (0.038) &  (0.011) \\
%\textit{Consuming Alcohol} & $0.289^{***}$ & $0.123^{***}$ & $0.168^{***}$ \\
%& (0.025) & (0.028) & (0.014) \\
%\midrule
AME of one friend shifting to the high action& {Homogeneous model} & \multicolumn{2}{c}{Heterogeneous model} \\ 
 \cmidrule(lr){1-1} \cmidrule(lr){2-2} \cmidrule(lr){3-4} 
\textit{Smoking} & $0.146^{***}$ & \multicolumn{2}{c}{$0.115^{***}$} \\
& (0.021) & \multicolumn{2}{c}{(0.004)} \\
\textit{Consuming Alcohol} & $0.154^{***}$ & \multicolumn{2}{c}{$0.154^{***}$} \\
& (0.011) & \multicolumn{2}{c}{(0.006)}\\
\bottomrule
\end{tabular}
\begin{tablenotes} 
\item \footnotesize \textit{Note}: $^{*} p-value < 0.05$, $^{**} p-value < 0.01$, $^{***} p-value< 0.001$. The terms in brackets represent the standard errors, calculated using the Delta method. AME of one friend shifting to the high action indicates the average marginal effect of expecting one friend to shift from the low to the high action. 
\end{tablenotes}
\caption{Average marginal effects of the social norm for non-isolated students}
\label{margin1}
\end{threeparttable}
\end{center}
\end{comment}
\noindent The marginal effects of the norm are of the standard logit form, $\frac{\partial p^*_i}{\partial \bar{p}_i^*}= \mathds{1}\{d_i>0\}\left(\hat{\beta^h}+\widehat{\Delta \beta}\bar{p}_i^*\right)p^*_i(1-p^*_i)$. The sample average marginal effect of an increase in the social norm is strong: for non-isolated students, expecting a friend to shift from the low to the high action increases the probability of selecting the high action in the heterogeneous model by 0.115pp and 0.154pp for smoking and alcohol consumption, respectively.\footnote{Non-isolated students have on average 3.84 friends, such that expecting one friend to shift to the high action corresponds to an increase of 0.26pp of the norm.} In the homogeneous model, the average marginal effect of the social norm is weakly lower, reaching 0.146pp and 0.154pp for tobacco and alcohol consumption, respectively. Peer effects are thus underestimated in the homogeneous specification in the case of smoking. \\

\subsection{Policy Simulations}
In this section, I simulate the effects of several policies that a social planner could implement to reduce tobacco and alcohol consumption among U.S. high school students. I focus on nationwide policies, such as advertising campaigns or social nudges, that require only the estimation of conformity preferences using a representative sample of the student population, as reported in Table \ref{estimation}. Accordingly, I do not consider more costly interventions that necessitate detailed network data of the entire student population, such as subsidies designed to restore first-best outcomes, as in \cite{boucher2023}, or key-player interventions, as in \cite{galeotti2020} and \cite{ballester2006}. The social planner's objective is to achieve a \textit{lower} equilibrium by minimizing the following objective function:  
$$\mathcal{V}(p^*)=\sum_i^n p_i^*.$$
As shown in Corollary \ref{compa}$(i)$ and $(ii)$, to reduce smoking and drinking, the social planner can increase the taste for conformity when choosing the high action, $\beta^h$, or reduce the cost of choosing the low action, $\beta^l$. Social marketing campaigns that emphasize how smoking or consuming alcohol can lead to social ostracization, or that reassure students that not smoking or drinking does not entail social penalties, could help achieve this.
Corollary \ref{compa}$(iii)$ shows that, since $\beta^h, \beta^l \geq 0$, the planner can also shift students’ perceptions of the idiosyncratic norm, $\bar{p}_i$, by $10\%$ toward zero to reduce the prevalence of smoking and drinking. This could be achieved by convincing students that they tend to overestimate the share of smokers or alcohol consumers by $10\%$. I thus evaluate how students would best-respond to $0.9p\bar p_i$, treating it as an exogenous perceived norm rather than an equilibrium outcome. \\

\begin{table}[h!]
\centering
\begin{threeparttable}
\begin{tabular}{lllrr}
\toprule
\multirow{2}{*}{Policy} & \multirow{2}{*}{Model} & \multirow{2}{*}{Scenario} & Smoking & Drinking \\
& & & $\Delta \mathcal{V}(\mathbf p^*)$ & $\Delta \mathcal{V}(\mathbf p^*)$ \\
\midrule
\multirow{2}{*}{Ad. Campaign} & \multirow{2}{*}{Homogeneous} 
& $1.1\beta$ & -2.64\% & 0.23\% \\
& & $0.9 \beta$ & 2.77\% & -0.23\% \\
\midrule
\multirow{4}{*}{Ad. Campaign} & \multirow{4}{*}{Heterogeneous} 
& $1.1 \beta^h$ & -1.38\% & -0.57\% \\
&& $1.1\beta^h$, $1.1\beta^l$  & -0.13\% & 0.51\% \\
& & $0.9 \beta^l$ & -1.26\% & -1.10\% \\
& & $0.9 \beta^h$, $0.9 \beta^l$& 0.15\%  & -0.52\% \\
\midrule
Nudge & Homogeneous & $0.9 \bar{p}_i$ & -3.08\% & -2.83\% \\
\midrule
Nudge & Heterogeneous & $0.9 \bar{p}_i$ & -3.31\% & -3.30\% \\
\bottomrule
\end{tabular}

\begin{tablenotes}
\footnotesize
\item \textit{Notes:} $\Delta \mathcal{V}(\mathbf p^*)$ represents the percentage change in $\mathcal{V}(\mathbf p^*) = \sum_{i=1}^n p_i^*$ relative to the baseline equilibrium without policy intervention. Negative values indicate a reduction in smoking or drinking prevalence.
\end{tablenotes}

\end{threeparttable}
\caption{Simulated effects of policies on the prevalence of smoking and drinking}\label{simu}
\end{table}

The simulated effects of social marketing campaigns differ substantially between the homogeneous and heterogeneous conformity models (Table~\ref{simu}). While an increase in $\beta$ in the homogeneous model decreases smoking prevalence, the equivalent policy in the heterogeneous model, a joint increase in $\beta^h$ and $\beta^l$, is about 20 times less effective. A planner relying on the homogeneous specification would therefore overestimate the policy’s effectiveness in an ex-ante evaluation. In contrast, targeted policies that either increase $\beta^h$ or decrease $\beta^l$ produce substantial reductions in smoking.
Similar patterns hold for alcohol consumption, except that the planner should decrease $\beta$ to reduce drinking prevalence, since a majority (55.4\%) of students consume alcohol. A joint decrease in $\beta^h$ and $\beta^l$ also reduces alcohol consumption, but to a lesser extent than a targeted policy that decreases $\beta^l$ alone. In addition, if the planner considers a homogeneous model, she would underestimate the impact of the social marketing campaign.

A social nudge is 7.5\% more effective under the heterogeneous model than predicted by the homogeneous model for smoking, and 16.6\% more effective for drinking. This reflects the higher taste for conformity when choosing the high action for both behaviors: the social nudge lowers the expected norm, thereby increasing the utility of not smoking or drinking.

Although the bias in the ex-ante evaluation of social nudges is substantial, a social planner using the homogeneous conformity specification would still correctly conclude that this policy is effective. By contrast, the bias in evaluating social marketing campaigns is more consequential. If the planner assumes homogeneous conformity preferences and shifts both conformity parameters uniformly (e.g., a 10\% decrease in both $\beta^h$ and $\beta^l$ for drinking, or a 10\% increase for smoking), the true effects could be 2 times larger or 20 times smaller, respectively, than what she would estimate. In addition, for both smoking and alcohol consumption, recognizing the heterogeneity in conformity preferences allows the planner to target only the relevant parameter (decreasing $\beta^l$ or increasing 
$\beta^h$) and be more effective than with a joint targeting of both parameters.

\section{Conclusion}

\noindent This paper introduces a structural conformity model that allows for action-specific heterogeneity in the preference for conformity to the social norm. The microfoundations of this model are established within a network game of incomplete information and a binary action space. %I consider an incomplete information setting where players do not observe their peers' actions but form rational expectations about them.
The Bayes-Nash equilibrium is unique under a mild assumption on the strength of the heterogeneous tastes for conformity. I show that the model's parameters are identified, even with heterogeneous conformity preferences. I also propose a specification test to infer whether the conformity or spillover models are consistent with the data, when action-specific heterogeneity and isolated players are present.% and potential misclassification of the outcome. Correcting for misclassification is particularly important empirically since it leads to a misassignment of the conformity parameters in the utility function.  

Using the NPL estimator of \cite{aguirregabiria2007}, I estimate the heterogeneous model on smoking behavior among US high school students and illustrate how assuming a homogeneous conformity preference, as in \cite{lee_binary_2014}, may lead to biased estimates of endogenous peer effects. I uncover the intuitive result that deviating from the norm is more costly for non-smokers than smokers, because of passive smoking and the addictive nature of smoking. For alcohol consumption, I find that the taste for conformity is similar when choosing to drink or not. I can therefore run the specification test only for smoking, which rejects the null hypothesis that the spillover model is consistent with the data but cannot reject the null for the conformity model below the 0.2\% significance level. Simulation exercises demonstrate that relying on a homogeneous conformity model may lead to severely biased ex-ante evaluations of public policies. 

Based on these empirical findings, I encourage further research efforts to incorporate action-specific heterogeneity in peer effects models for count, multinomial and continuous outcomes. %and to develop identification and estimation methods that are robust to mismeasurement in social interaction models. 

\clearpage
\newpage

\bibliographystyle{agsm}
\bibliography{bibliopeers2}

@article{nakajima2007,
  title={Measuring peer effects on youth smoking behaviour},
  author={Nakajima, Ryo},
  journal={The Review of Economic Studies},
  volume={74},
  number={3},
  pages={897--935},
  year={2007},
  publisher={Wiley-Blackwell}
}

@misc{houndetoungan2025,
      title={Quantile Peer Effect Models}, 
      author={Aristide Houndetoungan},
      year={2025},
      eprint={2506.12920},
      archivePrefix={arXiv},
      primaryClass={econ.EM},
      url={https://arxiv.org/abs/2506.12920}, 
}

@article{menzel2016,
  title={Inference for games with many players},
  author={Menzel, Konrad},
  journal={The Review of Economic Studies},
  volume={83},
  number={1},
  pages={306--337},
  year={2016},
  publisher={Oxford University Press}
}

@article{rust1987,
  title={Optimal replacement of GMC bus engines: An empirical model of Harold Zurcher},
  author={Rust, John},
  journal={Econometrica: Journal of the Econometric Society},
  pages={999--1033},
  year={1987},
  publisher={JSTOR}
}

@article{manski1988,
  title={Identification of binary response models},
  author={Manski, Charles F},
  journal={Journal of the American statistical Association},
  volume={83},
  number={403},
  pages={729--738},
  year={1988},
  publisher={Taylor \& Francis}
}

@article{brock2007,
  title={Identification of binary choice models with social interactions},
  author={Brock, William A and Durlauf, Steven N},
  journal={Journal of Econometrics},
  volume={140},
  number={1},
  pages={52--75},
  year={2007},
  publisher={Elsevier}
}

@article{lin2021,
  title={Uncovering heterogeneous social effects in binary choices},
  author={Lin, Zhongjian and Tang, Xun and Yu, Ning Neil},
  journal={Journal of Econometrics},
  volume={222},
  number={2},
  pages={959--973},
  year={2021},
  publisher={Elsevier}
}

@article{ciliberto2009,
  title={Market structure and multiple equilibria in airline markets},
  author={Ciliberto, Federico and Tamer, Elie},
  journal={Econometrica},
  volume={77},
  number={6},
  pages={1791--1828},
  year={2009},
  publisher={Wiley Online Library}
}

@article{zenou2025,
  title={Peer vs. Network Effects: Microfoundations, Identification, and Beyond},
  author={Zenou, Yves},
  journal={SSRN},
  year={2025}
}

@article{guerra2022,
  title={Multinomial choice with social interactions: Occupations in Victorian London},
  author={Guerra, Jos{\'e}-Alberto and Mohnen, Myra},
  journal={Review of Economics and Statistics},
  volume={104},
  number={4},
  pages={736--747},
  year={2022},
  publisher={MIT Press One Rogers Street, Cambridge, MA 02142-1209, USA journals-info~…}
}

@article{krauth2006,
  title={Simulation-based estimation of peer effects},
  author={Krauth, Brian V},
  journal={Journal of Econometrics},
  volume={133},
  number={1},
  pages={243--271},
  year={2006},
  publisher={Elsevier}
}

@article{tamer2003,
  title={Incomplete simultaneous discrete response model with multiple equilibria},
  author={Tamer, Elie},
  journal={The Review of Economic Studies},
  volume={70},
  number={1},
  pages={147--165},
  year={2003},
  publisher={Wiley-Blackwell}
}

@article{aguirregabiria2007,
  title={Sequential estimation of dynamic discrete games},
  author={Aguirregabiria, Victor and Mira, Pedro},
  journal={Econometrica},
  volume={75},
  number={1},
  pages={1--53},
  year={2007},
  publisher={Wiley Online Library}
}

@article{soetevent2007,
  title={A discrete-choice model with social interactions: With an application to high school teen behavior},
  author={Soetevent, Adriaan R and Kooreman, Peter},
  journal={Journal of Applied Econometrics},
  volume={22},
  number={3},
  pages={599--624},
  year={2007},
  publisher={Wiley Online Library}
}

@article{akerlof1997,
	title = {Social Distance and Social Decisions},
	volume = {65},
	doi = {10.2307/2171877},
	language = {en},
	number = {95},
		journal = {Econometrica},
	author = {Akerlof, George},
	year = {1997},
	pages = {1005},
	
}

@article{rothenberg1971,
  title={Identification in parametric models},
  author={Rothenberg, Thomas J},
  journal={Econometrica: Journal of the Econometric Society},
  pages={577--591},
  year={1971},
  publisher={JSTOR}
}

@incollection{jackson2015,
  title={Games on networks},
  author={Jackson, Matthew O and Zenou, Yves},
  booktitle={Handbook of game theory with economic applications},
  volume={4},
  pages={95--163},
  year={2015},
  publisher={Elsevier}
}

@article{Bramoulle2009,
	title = {Identification of peer effects through social networks},
	volume = {150},
	journal = {Journal of Econometrics},
	author = {Bramoullé, Y. and Djebbari, H. and Fortin, B.},
	year = {2009},
	pages = {41--55},
}

@article{ballester2006,
  title={Who's who in networks. Wanted: The key player},
  author={Ballester, Coralio and Calv{\'o}-Armengol, Antoni and Zenou, Yves},
  journal={Econometrica},
  volume={74},
  number={5},
  pages={1403--1417},
  year={2006},
  publisher={Wiley Online Library}
}

@article{lambotte2024,
  title={Peer Effects in Binary Outcomes: Strategic Complementarity and Taste for Conformity with Endogenous Networks},
  author={Lambotte, Mathieu},
  journal={Journal of Applied Econometrics},
  year={2025}
}

@article{bernheim1984,
  title={Rationalizable strategic behavior},
  author={Bernheim, B Douglas},
  journal={Econometrica: Journal of the Econometric Society},
  pages={1007--1028},
  year={1984},
  publisher={JSTOR}
}

@article{lin2024,
  title={Quantile Effects in Discrete Choice with Social Interactions},
  author={Lin, Zhongjian},
  journal={Forthcoming at The Review of Economics and Statistics},
  year={2024}
}

@article{liu2019,
  title={Simultaneous equations with binary outcomes and social interactions},
  author={Liu, Xiaodong},
  journal={Econometric Reviews},
  volume={38},
  number={8},
  pages={921--937},
  year={2019},
  publisher={Taylor \& Francis}
}

@article{lin2017,
  title={Estimation of social-influence-dependent peer pressure in a large network game},
  author={Lin, Zhongjian and Xu, Haiqing},
  journal={The Econometrics Journal},
  volume={20},
  number={3},
  pages={S86--S102},
  year={2017},
  publisher={Oxford University Press Oxford, UK}
}

@article{liu2017,
  title={A social interaction model with ordered choices},
  author={Liu, Xiaodong and Zhou, Jiannan},
  journal={Economics Letters},
  volume={161},
  pages={86--89},
  year={2017},
  publisher={Elsevier}
}

@article{lee_identification_2007,
	title = {Identification and {Estimation} of {Econometrics} {Models} with {Group} interactions, contextual factors and fixed effects},
	volume = {140},
	number = {2},
	journal = {Journal of Econometrics},
	author = {Lee, L. F.},
	year = {2007},
	pages = {333--374},
}

@article{hsieh2020,
  title={Specification and estimation of network formation and network interaction models with the exponential probability distribution},
  author={Hsieh, Chih-Sheng and Lee, Lung-Fei and Boucher, Vincent},
  journal={Quantitative economics},
  volume={11},
  number={4},
  pages={1349--1390},
  year={2020},
  publisher={Wiley Online Library}
}

@article{houndetoungan_count_2020,
  title={Count Data Models with Heterogeneous Peer Effects under Rational Expectations},
  author={Houndetoungan, Aristide},
  journal={Journal of Applied Econometrics},
  year={2026},
  publisher={Wiley Online Library}
}

@article{graham2005,
  title={Identification and estimation of the linear-in-means model of social interactions},
  author={Graham, Bryan S and Hahn, Jinyong},
  journal={Economics Letters},
  volume={88},
  number={1},
  pages={1--6},
  year={2005},
  publisher={Elsevier}
}

@article{watts1998,
  title={Collective dynamics of ‘small-world’networks},
  author={Watts, Duncan J and Strogatz, Steven H},
  journal={nature},
  volume={393},
  number={6684},
  pages={440--442},
  year={1998},
  publisher={Nature Publishing Group}
}

@article{jackson2007,
  title={Meeting strangers and friends of friends: How random are social networks?},
  author={Jackson, Matthew O and Rogers, Brian W},
  journal={American Economic Review},
  volume={97},
  number={3},
  pages={890--915},
  year={2007},
  publisher={American Economic Association}
}

@article{schultz2007,
  title={The constructive, destructive, and reconstructive power of social norms},
  author={Schultz, P Wesley and Nolan, Jessica M and Cialdini, Robert B and Goldstein, Noah J and Griskevicius, Vladas},
  journal={Psychological science},
  volume={18},
  number={5},
  pages={429--434},
  year={2007},
  publisher={SAGE Publications Sage CA: Los Angeles, CA}
}

@article{davezies2009,
  title={Identification of peer effects using group size variation},
  author={Davezies, Laurent and d'Haultfoeuille, Xavier and Foug{\`e}re, Denis},
  journal={The Econometrics Journal},
  volume={12},
  number={3},
  pages={397--413},
  year={2009},
  publisher={Oxford University Press Oxford, UK}
}

@article{leung2015,
  title={Two-step estimation of network-formation models with incomplete information},
  author={Leung, Michael P},
  journal={Journal of Econometrics},
  volume={188},
  number={1},
  pages={182--195},
  year={2015},
  publisher={Elsevier}
}

@article{horst2006,
  title={Equilibria in systems of social interactions},
  author={Horst, Ulrich and Scheinkman, Jose A},
  journal={Journal of Economic Theory},
  volume={130},
  number={1},
  pages={44--77},
  year={2006},
  publisher={Elsevier}
}

@article{boucheraristide,
  title={Estimating peer effects using partial network data},
  author={Boucher, Vincent and Houndetoungan, Aristide},
  year={2025}
}

@inbook{glaeser2003, place={Cambridge}, series={Econometric Society Monographs}, title={Nonmarket Interactions}, booktitle={Advances in Economics and Econometrics: Theory and Applications, Eighth World Congress}, publisher={Cambridge University Press}, author={Glaeser, Edward and Scheinkman, José A.}, year={2003}, pages={339–370}, collection={Econometric Society Monographs}}

@article{leung2020,
  title={Equilibrium computation in discrete network games},
  author={Leung, Michael P},
  journal={Quantitative Economics},
  volume={11},
  number={4},
  pages={1325--1347},
  year={2020},
  publisher={Wiley Online Library}
}

@article{li2016,
  title={A partial identification subnetwork approach to discrete games in large networks: An application to quantifying peer effects},
  author={Li, Tong and Zhao, Li},
  journal={The Institute for Fiscal Studies},
  year={2016}
}

@article{depaula2013,
  title={Econometric analysis of games with multiple equilibria},
  author={De Paula, Aureo},
  journal={Annu. Rev. Econ.},
  volume={5},
  number={1},
  pages={107--131},
  year={2013},
  publisher={Annual Reviews}
}

@article{lopez2021,
  title={Far above others},
  author={L{\'o}pez-Pintado, Dunia and Mel{\'e}ndez-Jim{\'e}nez, Miguel A},
  journal={Journal of Economic Theory},
  volume={198},
  pages={105376},
  year={2021},
  publisher={Elsevier}
}

@article{immorlica2017,
  title={Social status in networks},
  author={Immorlica, Nicole and Kranton, Rachel and Manea, Mihai and Stoddard, Greg},
  journal={American Economic Journal: Microeconomics},
  volume={9},
  number={1},
  pages={1--30},
  year={2017},
  publisher={American Economic Association 2014 Broadway, Suite 305, Nashville, TN 37203-2425}
}

@article{azmat2010,
  title={The importance of relative performance feedback information: Evidence from a natural experiment using high school students},
  author={Azmat, Ghazala and Iriberri, Nagore},
  journal={Journal of Public Economics},
  volume={94},
  number={7-8},
  pages={435--452},
  year={2010},
  publisher={Elsevier}
}

@article{fehr1999,
  title={A theory of fairness, competition, and cooperation},
  author={Fehr, Ernst and Schmidt, Klaus M},
  journal={The quarterly journal of economics},
  volume={114},
  number={3},
  pages={817--868},
  year={1999},
  publisher={MIT press}
}

@article{galeotti2020,
  title={Targeting interventions in networks},
  author={Galeotti, Andrea and Golub, Benjamin and Goyal, Sanjeev},
  journal={Econometrica},
  volume={88},
  number={6},
  pages={2445--2471},
  year={2020},
  publisher={Wiley Online Library}
}

@article{fershtman1998,
  title={Social rewards, externalities and stable preferences},
  author={Fershtman, Chaim and Weiss, Yoram},
  journal={Journal of Public Economics},
  volume={70},
  number={1},
  pages={53--73},
  year={1998},
  publisher={Elsevier}
}

@article{kandel1992,
  title={Peer pressure and partnerships},
  author={Kandel, Eugene and Lazear, Edward P},
  journal={Journal of Political Economy},
  volume={100},
  number={4},
  pages={801--817},
  year={1992},
  publisher={The University of Chicago Press}
}

@article{akerlof1980,
  title={A theory of social custom, of which unemployment may be one consequence},
  author={Akerlof, George A},
  journal={The quarterly journal of economics},
  volume={94},
  number={4},
  pages={749--775},
  year={1980},
  publisher={MIT Press}
}

@article{patacchini2014,
  title={Peer effects in the demand for housing quality},
  author={Patacchini, Eleonora and Venanzoni, Giuseppe},
  journal={Journal of Urban Economics},
  volume={83},
  pages={6--17},
  year={2014},
  publisher={Elsevier}
}

@article{kasahara2012,
  title={Sequential estimation of structural models with a fixed point constraint},
  author={Kasahara, Hiroyuki and Shimotsu, Katsumi},
  journal={Econometrica},
  volume={80},
  number={5},
  pages={2303--2319},
  year={2012},
}

@article{chandrasekhar2011,
  title={Econometrics of sampled networks},
  author={Chandrasekhar, Arun and Lewis, Randall},
  journal={Unpublished manuscript, MIT.[422]},
  volume={2},
  pages={7},
  year={2011}
}

@article{boucher2023,
  title={Toward a general theory of peer effects},
  author={Boucher, Vincent and Rendall, Michelle and Ushchev, Philip and Zenou, Yves},
  year={2024},
  journal={Econometrica},
volume={92},
issue={2},
  pages={543--565},
}

@article{blume_linear_2015,
	title = {Linear {Social} {Interactions} {Models}},
	volume = {123},
	issn = {0022-3808, 1537-534X},
	doi = {10.1086/679496},
	language = {en},
	number = {2},
	urldate = {2022-01-25},
	journal = {Journal of Political Economy},
	author = {Blume, Lawrence E. and Brock, William A. and Durlauf, Steven N. and Jayaraman, Rajshri},
	month = apr,
	year = {2015},
	pages = {444--496},
	
}

@article{xu2018,
  title={Social interactions in large networks: A game theoretic approach},
  author={Xu, Haiqing},
  journal={International Economic Review},
  volume={59},
  number={1},
  pages={257--284},
  year={2018},
  publisher={Wiley Online Library}
}

@article{ushchev2020,
  title={Social norms in networks},
  author={Ushchev, Philip and Zenou, Yves},
  journal={Journal of Economic Theory},
  volume={185},
  pages={104969},
  year={2020},
  publisher={Elsevier}
}

@article{lee_binary_2014,
	title = {Binary {Choice} {Models} with {Social} {Network} under {Heterogeneous} {Rational} {Expectations}},
	volume = {96},
		number = {3},
	urldate = {2022-01-25},
	journal = {Review of Economics and Statistics},
	author = {Lee, Lung-fei and Li, Ji and Lin, Xu},
		year = {2014},
	pages = {402--417},
	}

@article{Badev2021,
author = {Badev, Anton},
title = {Nash Equilibria on (Un)Stable Networks},
journal = {Econometrica},
volume = {89},
number = {3},
pages = {1179-1206},
doi = {https://doi.org/10.3982/ECTA12576},
year = {2021}
}

@article{Boucher2016,
title = {Conformism and self-selection in social networks},
journal = {Journal of Public Economics},
volume = {136},
pages = {30-44},
year = {2016},
issn = {0047-2727},
doi = {https://doi.org/10.1016/j.jpubeco.2016.02.005},
author = {Vincent Boucher},
}

@article{Bramoulle2019,
	title = {Peer Effects in Networks: a Survey},
	language = {en},
	author = {Bramoullé, Yann and Djebbari, Habiba and Fortin, Bernard},
	pages = {603--619},
	volume = {12},
	journal = {Annual Review of Economics},
	year = {2019},
}

@article{li2009,
  title={Binary choice under social interactions: an empirical study with and without subjective data on expectations},
  author={Li, Ji and Lee, Lung-fei},
  journal={Journal of Applied Econometrics},
  volume={24},
  number={2},
  pages={257--281},
  year={2009},
  publisher={Wiley Online Library}
}

@article{bajari2010identification,
  title={Identification and estimation of a discrete game of complete information},
  author={Bajari, Patrick and Hong, Han and Ryan, Stephen P},
  journal={Econometrica},
  volume={78},
  number={5},
  pages={1529--1568},
  year={2010},
  publisher={Wiley Online Library}
}

@article{brock_discrete_2001,
	title = {Discrete {Choice} with {Social} {Interactions}},
	volume = {68},
	issn = {0034-6527, 1467-937X},
	doi = {10.1111/1467-937X.00168},
	language = {en},
	number = {2},
	urldate = {2022-01-25},
	journal = {The Review of Economic Studies},
	author = {Brock, W. A. and Durlauf, S. N.},
	month = apr,
	year = {2001},
	pages = {235--260},

}

@article{bhattacharya2024,
  title={Demand and welfare analysis in discrete choice models with social interactions},
  author={Bhattacharya, Debopam and Dupas, Pascaline and Kanaya, Shin},
  journal={Review of Economic Studies},
  volume={91},
  number={2},
  pages={748--784},
  year={2024},
  publisher={Oxford University Press US}
}

@article{hsieh2018,
  title={Smoking initiation: Peers and personality},
  author={Hsieh, Chih-Sheng and Van Kippersluis, Hans},
  journal={Quantitative Economics},
  volume={9},
  number={2},
  pages={825--863},
  year={2018},
  publisher={Wiley Online Library}
}

\clearpage
\newpage

\appendix
\renewcommand{\thesection}{\Alph{section}}
\begin{center}
    \Large \bf Appendix
\end{center}
\section{Proof of Proposition \ref{P1}} \label{AppendixB}
I prove that $\Gamma(\mathbf{p})$ is a contraction mapping for the metric 
$d(\mathbf{p},\mathbf{q})=\left\| \mathbf{p}-\mathbf{q}\right\|_{\infty}$, 
where $\|\mathbf{X}\|_{\infty}=\max_i\sum_j \lvert x_{ij} \rvert$ for any matrix $\mathbf{X}$. 
By the mean-value theorem, it suffices to show that 
$\left\|\frac{\partial \Gamma(\mathbf{p})}{\partial \mathbf{p}}\right\|_\infty < 1$.\\

\noindent Let $\mathbf{f}_m^* = (f_1^*,\ldots,f_{n_m}^*)'$ where 
$f_i^* = f_\eta\left(\boldsymbol\alpha_i+\beta^h\left(\bar{p}_i^*-\frac{1}{2}\right)+\frac{1}{2}\Delta \beta\mathbf g_i\boldsymbol\Sigma \mathbf g_i^{\prime})\right)$, where $f_\eta$ is the density of $F_\eta$. 
Recall that $g_i\boldsymbol\Sigma \mathbf g_i^{\prime}=\mathbf{g}_i\left(\mathbf{p} \mathbf p^{\prime}+\operatorname{diag}(\mathbf p\circ(\mathbf{1}_n-\mathbf p))\right)\mathbf{g}^{\prime}_i$, which can be written as $(\mathbf g_i \mathbf p)^2+\mathbf{g}_i\left(\operatorname{diag}(\mathbf p\circ(\mathbf{1}_n-\mathbf p))\right)\mathbf{g}^{\prime}_i$. Let $\mathbf e_j$ be the $j^{th}$ standard basis vector. The derivative of $g_i\boldsymbol\Sigma \mathbf g_i^{\prime}$ with respect to $p_j$ is given by:
\begin{align*}
    \frac{\partial g_i\boldsymbol\Sigma \mathbf g_i^{\prime}}{\partial p_j}&=2(\mathbf{g}_i \mathbf{e}_j')(\mathbf{g}_i \mathbf{p}) + \mathbf{g}_i(1-2p_j)\mathbf{e}_j \mathbf{e}_j' \mathbf{g}_i\\
    &=2g_{ij}\bar{p}_i + g_{ij}^2(1-2p_j)
\end{align*}

\noindent The partial derivative of the best-response mapping is then given by:
\begin{equation*}
    \frac{\partial \Gamma_i(\mathbf{p})}{\partial p_j} 
    = \left(\beta^h g_{ij} + \Delta\beta g_{ij}\bar{p}_i  
    + \frac{1}{2}\Delta\beta\, g_{ij}^2(1-2p_j)\right) f_i^*
\end{equation*}
 Define $c_{ij} = \beta^h g_{ij} + \Delta\beta g_{ij}\bar{p}_i 
    + \frac{1}{2}\Delta\beta\, g_{ij}^2(1-2p_j)$, $\mathbf{c}_i=(c_{i1},\cdots,c_{in})$ and $\mathbf{C} = [\mathbf c_{1}' \cdots \mathbf c_{n}']'$.
The Jacobian of $\Gamma(\mathbf{p})$ can then be written compactly as:
\begin{equation}\label{eqA.1}
    \frac{\partial \Gamma(\mathbf{p})}{\partial \mathbf{p}} 
    = \operatorname{diag}(\mathbf{f}^*)\,\mathbf{C} \tag{A.1}
\end{equation}
Taking the $L_\infty$ norm:
\begin{align*}
\left\|\frac{\partial \Gamma(\mathbf{p})}{\partial \mathbf{p}}\right\|_\infty
&= \left\|\operatorname{diag}(\mathbf{f}^*)\,\mathbf{C}\right\|_\infty \\
&= \max_{i\in\mathcal{N}} f_i^* \sum_{j=1}^{n} \lvert c_{ij} \rvert \\
&\leq \max_{i\in\mathcal{N}} f_i^* \sum_{j=1}^{n} 
   \left(\lvert \beta^h \rvert g_{ij} + \lvert \Delta\beta \rvert \bar{p}_i g_{ij} 
   + \frac{1}{2}\lvert \Delta\beta\rvert g_{ij}^2\lvert 1-2p_j\rvert \right) \\
&\leq \max_u f_\eta(u)
   \left(\lvert \beta^h \rvert + \lvert \Delta\beta \rvert  + \frac{1}{2}\lvert \Delta\beta \rvert \right) \\
&= \max_u f_\eta(u)\left(\lvert \beta^h \rvert+\frac{3}{2}\lvert \beta^l-\beta^h \rvert\right)
\end{align*}
where the second equality follows from the definition of the $L_\infty$ norm and because $f_i^*>0$, and the first inequality uses the triangle inequality. 
The second inequality uses $g_{ij}\in[0,1]$, $\bar{p}_i\in[0,1]$, 
$\lvert 1-2p_j\rvert \leq 1$, and:
\begin{equation*}
    \max_{i\in\mathcal{N}}\sum_j g_{ij}^2\,\lvert 1-2p_j\rvert  
    \leq \max_{i\in\mathcal{N}}\sum_j g_{ij}^2 
    \leq \max_{i\in\mathcal{N}}\sum_j g_{ij} 
    = \|\mathbf{G}_m\|_\infty = 1,
\end{equation*}
where the last equality follows from row-normalization of $\mathbf{G}_m$.

It follows that $\left\|\frac{\partial \Gamma( \mathbf p)}{\partial \mathbf p}\right\|_\infty<1$ if and only if: 
 \begin{align*}\label{eqA.2}
 &   \lvert \beta^h \rvert+\frac{3}{2} \lvert \beta^l-\beta^h \rvert \max_u f_{\eta}(u) < 1 \\
 &\iff \lvert \beta^h \rvert+\frac{3}{2} \lvert \beta^l-\beta^h \rvert <\frac{1}{\max_u f_{ \eta}(u)} \tag{A.2}
 \end{align*}
 
\noindent If Equation (\ref{eqA.2}) holds, then from the contraction mapping theorem stated above, $\Gamma$ is a contraction mapping and the equilibrium $\mathbf{p}^*$ is unique. \hfill $\blacksquare$

\newpage
\section{Proof of Proposition \ref{P2}} \label{AppendixC}

Let $\boldsymbol \theta$ and $\tilde{\boldsymbol \theta}$ be two sets of parameters such that
\begin{align*}
\mathbf p_m &= F_\eta\left(\mathbf{Z}_m,\mathbf{G}_m;\boldsymbol\theta\right), \\
\tilde{\mathbf p}_m &= F_\eta\left(\mathbf{Z}_m,\mathbf{G}_m;\boldsymbol{\tilde{\theta}}\right).
\end{align*} 
The model is identified if there does not exist a $\tilde{\boldsymbol \theta} \ne \boldsymbol \theta$ that is observationally equivalent to $\boldsymbol \theta$, i.e., such that $\mathbf p_m = \tilde{\mathbf p}_m$. Observational equivalence then implies
\[
F_\eta\left(\mathbf{Z}_m,\mathbf{G}_m;\boldsymbol\theta\right) = F_\eta\left(\mathbf{Z}_m,\mathbf{G}_m;\boldsymbol{\tilde{\theta}}\right),
\]
since Assumption \ref{H1}$(ii)$ guarantees that $F_{\eta}$ is strictly increasing. Using Equation (\ref{eq7}), the above equality can be written for each player $i$ in network $m$ as:
\begin{equation*} 
\begin{split}
\mathbf z^{\prime}_i\boldsymbol \gamma + \beta^h\left(\bar{p}_i - \frac{1}{2}\right) + \Delta\beta \frac{1}{2}\mathbf g_i \boldsymbol\Sigma \mathbf g_i^{\prime} 
&= \mathbf z^{\prime}_i \boldsymbol{\tilde{\gamma}} + \tilde{\beta}^h\left(\bar{p}_i - \frac{1}{2}\right) + \widetilde{\Delta\beta} \frac{1}{2} \mathbf g_i \boldsymbol\Sigma \mathbf g_i^{\prime} \\
&\iff \mathbf k_i^\prime \boldsymbol \theta = \mathbf k_i^\prime \boldsymbol{\tilde{\theta}}.
\end{split}
\end{equation*}
Hence, $\boldsymbol \theta$ and $\tilde{\boldsymbol \theta}$ are observationally equivalent only if
\[
\frac{1}{M}\sum_{m=1}^M \sum_{i=1}^{n_m} \mathbf{k}_i \mathbf{k}_i^\prime \boldsymbol{\theta} = \frac{1}{M}\sum_{m=1}^M \sum_{i=1}^{n_m} \mathbf{k}_i \mathbf{k}_i^\prime \tilde{\boldsymbol \theta}.
\]
Under Assumption \ref{H4}$(ii)$, the matrix $\operatorname{plim}_{M \to \infty} \frac{1}{M} \sum_{m=1}^M \sum_{i=1}^{n_m} \mathbf{k}_i \mathbf{k}_i^\prime$ is full rank, which implies that $\boldsymbol \theta$ and $\tilde{\boldsymbol \theta}$ are observationally equivalent only if $\boldsymbol \theta = \tilde{\boldsymbol \theta}$. \hfill $\blacksquare$
\newpage

\section{Proof of Proposition \ref{PropoAri}} \label{AppendixD}

I prove Proposition \ref{PropoAri} by contradiction, assuming $\beta^h, \beta^l\geq 0$. Let $\mathbf I_{n_m}$ be the $n_m$-dimensional identity matrix. For any $(i,j)\in \mathcal{N}_m$, the marginal effect of $x_{j,\kappa}$ on $p_i$ is given by the total differential of Equation \eqref{eq7}:
\begin{align*}\label{D.1}
   \frac{d p_i}{dx_{j,\kappa}}&=f_i^*\times \left(\beta^h\mathbf{g}_{i,m}+\frac{1}{2}\Delta\beta\left( 2 \bar{p}_i\mathbf{g}_{i,m}+(\mathbf{1}_{n_m}-2\mathbf{p}_m)^\prime\operatorname{diag}(\mathbf{g}_{i,m})^2 \right)\right)\frac{d \mathbf{p}_m}{d x_{j,\kappa}} +f^*_i\gamma_{2,\kappa}g_{ij}\\
   &=f_i^*\mathbf{c}_{i,m}\left[\mathbf{I}_{n_m}-\operatorname{diag}(\mathbf f^*_m)\mathbf{C}_m\right]^{-1}e_j+f^*_i\gamma_{2,\kappa}g_{ij} \tag{C.1}
\end{align*}
where $f_i^* = f_\eta\left(\boldsymbol\alpha_i+\beta^h\left(\bar{p}_i^*-\frac{1}{2}\right)+\frac{1}{2}\Delta \beta\mathbf g_i\boldsymbol\Sigma \mathbf g_i^{\prime})\right)$, $\mathbf{e}_j = (e_{j,1}, \ldots, e_{j,n_m})^{\prime}$ with $e_{j,j} = \gamma_{1,\kappa} f_j^*,\; e_{j,k} = \gamma_{2,\kappa} g_{jk} f_k^* \; \text{for all } k \neq j$, $\mathbf{c}_{i,m}=\beta^h\mathbf{g}_{i,m}+\frac{1}{2}\Delta\beta\left( 2 \bar{p}_i\mathbf{g}_{i,m}+(\mathbf{1}_{n_m}-2\mathbf{p}_m)^\prime\operatorname{diag}(\mathbf{g}_{i,m})^2 \right)$ and $\mathbf{C}_m=\left[\mathbf{c}_{1,m}^\prime \cdots \mathbf{c}_{n_m,m}^\prime\right]^\prime$. Let $\mathcal{J}_{i,m} = \{j : g_{ij} > 0\}$ denote the set that contains all friends of $i$ in network $m$ and $\mathcal{K}_{ij,m} = \{k : g_{jk} > 0 \text{ and } g_{ik} = 0\}$ denote the set of $j$'s friends who are not $i$'s friends. Condition $(ii)$ in Proposition \ref{PropoAri} implies that $\mathcal{J}_{i,m}$ and $\mathcal{K}_{ij,m}$ are not empty for a positive number of agents $i$.\\

The following lemma guarantees that the marginal effect of $x_{j,\kappa}$ on $p_i$ is either negative or positive, for any peer $j$.
\begin{lemma}\label{Lemma1}
If $\beta^h, \beta^l\geq 0$, $\frac{d p_i}{dx_{j,\kappa}}$ is either strictly positive or strictly negative, for all $j\in \mathcal{J}_{i,m}$. For all $j \notin \mathcal{J}_{i,m}$, $\frac{d p_i}{dx_{j,\kappa}}= 0.$
\end{lemma}
\textbf{Proof:}
Because Condition $(iii)$ in Proposition \ref{PropoAri} imposes $\gamma_{1,\kappa}\gamma_{2,\kappa} \geq 0$, the elements of $\mathbf{e}_j$ and the term $f^*_i\gamma_{2,\kappa}g_{ij}$ have the same sign. If $\mathbf{c}_{i,m}\!\left[\mathbf{I}_{n_m}-\operatorname{diag}(\mathbf f^*_m)\mathbf{C}_m\right]^{-1}$ is non-negative, then 
$\frac{d p_i}{dx_{j,\kappa}}$ is strictly negative when $\gamma_{1,\kappa}$ is non-positive and $\gamma_{2,\kappa}$ is negative, and strictly positive when $\gamma_{1,\kappa}$ is non-negative and $\gamma_{2,\kappa}$ is positive. 

To establish that $\mathbf{c}_{i,m}\!\left[\mathbf{I}_{n_m}-\operatorname{diag}(\mathbf f^*_m)\mathbf{C}_m\right]^{-1}\geq 0$, it is sufficient to show that $\mathbf{C}_m$ is element-wise non-negative. Indeed, since $f_i^*\ge 0$ for all $i$, if $\mathbf{C}_m$ has non-negative elements then $\left[\mathbf{I}_{n_m}-\operatorname{diag}(\mathbf f^*_m)\mathbf{C}_m\right]^{-1}$, which is a Leontief inverse, has only non-negative elements. The Leontief inverse exists because the spectral radius satisfies $\rho\!\left(\operatorname{diag}(\mathbf f^*_m)\mathbf{C}_m\right)<1$, where $\rho(\cdot)$ denotes the spectral radius. To see this, recall that the proof of uniqueness of the equilibrium (Proposition \ref{P1}) has established $\big\|\operatorname{diag}(\mathbf f^*_m)\mathbf{C}_m\big\|_\infty<1$.
Since the spectral radius of a matrix is bounded above by any consistent matrix norm, it follows that $\rho\!\left(\operatorname{diag}(\mathbf f^*_m)\mathbf{C}_m\right) \leq \big\|\operatorname{diag}(\mathbf f^*_m)\mathbf{C}_m\big\|_\infty
<1$.\\

It remains to show that $c_{ij}\ge 0$ for all $(i,j)$, where $c_{ij}$ is given by:
\begin{equation*}
    c_{ij} = \beta^h g_{ij} + \Delta\beta\bar{p}_i g_{ij} + \frac{1}{2}\Delta\beta 
    g_{ij}^2(1-2p_j),
\end{equation*}
Dividing by $g_{ij} > 0$ for any $j\in\mathcal{J}_{i,m}$, $c_{ij}$ is non-negative if:
\begin{align*}
    &\beta^h + \Delta\beta \left(g_{ij}p_j+\sum\limits_{\substack{s \neq j, \\ s \in \mathcal{J}_{i,m}}}g_{is}p_s\right)+ \frac{1}{2}\Delta\beta g_{ij}(1-2p_j) \geq 0  \\
    & \iff \beta^h + \Delta\beta\sum\limits_{\substack{s \neq j, \\ s \in \mathcal{J}_{i,m}}}g_{is}p_s+ \frac{1}{2}\Delta\beta g_{ij} \geq0
\end{align*}
I distinguish two cases depending on the sign of $\Delta \beta$:
\begin{itemize}
    \item \textbf{$\Delta \beta \geq 0$}. Since $\sum\limits_{\substack{s \neq j, \\ s \in \mathcal{J}_{i,m}}}g_{is}p_s$ and $g_{ij}$ are positive, $c_{ij}\geq0 \iff \beta^h\geq 0$.
    \item \textbf{$\Delta \beta < 0$}. In that case, $ \Delta\beta\sum\limits_{\substack{s \neq j, \\ s \in \mathcal{J}_{i,m}}} g_{is}p_s$ is minimized when $p_s=1$, yielding $\sum\limits_{\substack{s \neq j, \\ s \in \mathcal{J}_{i,m}}}g_{is}p_s=\sum\limits_{\substack{s \neq j, \\ s \in \mathcal{J}_{i,m}}}g_{is}=1-g_{ij}$. In general, it follows that $ \Delta\beta\sum\limits_{\substack{s \neq j, \\ s \in \mathcal{J}_{i,m}}} g_{is}p_s\geq \Delta\beta(1-g_{ij})$ and thus that $\beta^h + \Delta\beta\sum\limits_{\substack{s \neq j, \\ s \in \mathcal{J}_{i,m}}}g_{is}p_s+ \frac{1}{2}\Delta\beta g_{ij}\geq \beta^h +\Delta\beta-\frac{1}{2}\Delta\beta g_{ij}$. Since $-\frac{1}{2}\Delta\beta g_{ij}\geq 0$, it follows that $c_{ij}\geq 0 \iff  \beta^h +\Delta\beta = \beta^l \geq 0$.
\end{itemize}
 Hence, $c_{ij}$ is non-negative for all $(i,j)$ if $\beta^h, \;\beta^l\geq 0$.  \hfill $\blacksquare$ \\

Then, observe that if Assumption \ref{H4}$(ii)$ does not hold, $\mathbf{k}_i$ is linearly dependent and, for any player $i$ and network $m$, I can write:
\begin{equation*}\label{D.2} \tag{C.2}
    \check{\beta}^h\left(\bar{p}_i-\frac{1}{2}\right)+\frac{1}{2}\widecheck{\Delta\beta}\mathbf \mathbf{g}_{i,m}\boldsymbol\Sigma_m \mathbf \mathbf{g}_{i,m}^{\prime}=\check{\gamma}_0 + \mathbf x_i^{\prime}\check{\boldsymbol\gamma}_1 +\bar{\mathbf{x}}_i^{\prime}\check{\boldsymbol\gamma}_2
\end{equation*}
for some constants $\check{\beta}^h\in \mathbb{R}_+$, $\widecheck{\Delta\beta}\in \mathbb{R}$, $\check{\gamma_0}\in \mathbb{R}$, $\check{\boldsymbol \gamma}_1\in \mathbb{R}^{d_\mathbf{X}}$ and $\check{\boldsymbol \gamma}_2\in \mathbb{R}^{d_\mathbf{X}}$. 

Since player $i$ has no friend in $\mathcal{K}_{ij,m} $, the right-hand side of Equation \eqref{D.2} does not depend on any player $k \in\mathcal{K}_{ij,m}$. Hence, by definition, the marginal effect of $x_{k,\kappa}$ on the left-hand side of Equation \eqref{D.2} must be zero:
\begin{equation*}\label{D.3}
    \frac{d\left[\check{\beta}^h\left(\bar{p}_i-\frac{1}{2}\right)+\frac{1}{2}\widecheck{\Delta\beta}\mathbf \mathbf{g}_{i,m}\boldsymbol\Sigma_m \mathbf \mathbf{g}_{i,m}^{\prime}\right]}{dx_{k,\kappa}}=\mathbf{\check{c}}_{i,m}\frac{d \mathbf{p}_m}{d x_{k,\kappa}}
    =0 \;\forall k \in \mathcal{K}_{ij,m}  \tag{C.3}
\end{equation*}
where $\mathbf{\check{c}}_{i,m}=\check{\beta}^h\mathbf{g}_{i,m}+\frac{1}{2}\widecheck{\Delta\beta}\left( 2 \bar{p}_i\mathbf{g}_{i,m}+(\mathbf{1}_{n_m}-2\mathbf{p}_m)^\prime\operatorname{diag}(\mathbf{g}_{i,m})^2 \right)$. Lemma \ref{Lemma1} shows that $\frac{d p_j}{d x_{k,\kappa}}\neq0$ and has the same sign for any $k \in \mathcal{K}_{ij,m} $ and that $\frac{d p_j}{d x_{k,\kappa}}\neq0$ for any $k \notin \mathcal{K}_{ij,m} $. Since $\mathcal{K}_{ij,m}$ is not empty for some $i$ in network $m$, Equation \eqref{D.3} holds only if $\mathbf{\check{c}}_{i,m}= 0$. 
In particular, the $j$-th element of $\mathbf{\check{c}}_{i,m}$ for any $j \in \mathcal{J}_{i,m}$ is:
\begin{equation*}\label{D.4}
  \left(\check{\beta}^h + \widecheck{\Delta\beta}\bar{p}_i\right) g_{ij} + \frac{\widecheck{\Delta\beta}}{2}g_{ij}^2(1-2p_j) = 0.\tag{C.4}
\end{equation*}

\noindent Differentiating Equation \eqref{D.4} with respect to $x_{k,\kappa}$ for any $k \in \mathcal{K}_{ij,m}$, i.e., a friend of $j$ who is not a friend of $i$, yields:
\begin{equation*}
    \widecheck{\Delta\beta}\, g_{ij}\frac{d\bar{p}_i}{dx_{k,\kappa}} 
    - \widecheck{\Delta\beta}\, g_{ij}^2\frac{dp_j}{dx_{k,\kappa}} = 0
\end{equation*}
Substituting $\frac{d\bar{p}_i}{dx_{k,\kappa}} = 
\sum_{\ell \in \mathcal{J}_{i,m}} g_{i\ell}\frac{dp_\ell}{dx_{k,\kappa}}$ and isolating the $\ell = j$ term, I obtain:
\begin{equation*}
    \widecheck{\Delta\beta}\, g_{ij}
    \sum_{\substack{\ell \in \mathcal{J}_{i,m} \cap \mathcal{J}_{k,m} 
    \\ \ell \neq j}} g_{i\ell}\frac{dp_\ell}{dx_{k,\kappa}}+\widecheck{\Delta\beta}\, g_{ij}^2\frac{dp_j}{dx_{k,\kappa}} 
        - \widecheck{\Delta\beta}\, g_{ij}^2\frac{dp_j}{dx_{k,\kappa}} = 0
\end{equation*}
The $\widecheck{\Delta\beta}\,g_{ij}^2\frac{dp_j}{dx_{k,\kappa}}$ terms cancel out, leaving:
\begin{equation} \label{C.5} \tag{C.5}
    \widecheck{\Delta\beta}\, g_{ij}
    \sum_{\substack{\ell \in \mathcal{J}_{i,m} \cap \mathcal{J}_{k,m} 
    \\ \ell \neq j}} g_{i\ell}\frac{dp_\ell}{dx_{k,\kappa}} = 0
\end{equation}
By Lemma~\ref{Lemma1}, all terms $\frac{dp_\ell}{dx_{k,\kappa}}$ for $\ell \in \mathcal{J}_{k,m}$ share the same sign, and $g_{i\ell}>0$ for all $\ell \in \mathcal{J}_{i,m}$, so no cancellation is possible within the sum. Therefore, Equation \eqref{C.5} holds only if $\widecheck{\Delta\beta} = 0$, provided that $\mathcal{J}_{i,m} \cap \mathcal{J}_{k,m} \setminus \{j\} \neq \emptyset$, i.e., $i$ and $k$ share at least one common friend other than $j$, which holds under Condition $(ii)$ in Proposition \ref{PropoAri}. Hence, Equation \eqref{D.4} simplifies to $\check{\beta}^h g_{ij}  = 0$. Since $g_{ij}>0$, it follows that $\check{\beta}^h=0$. \\

\noindent Since $\check{\beta}^h=0$ and $\widecheck{\Delta\beta} = 0$, Equation \eqref{D.2} simplifies to 
$\check{\gamma}_0 + \mathbf{x}_i^{\prime}\check{\boldsymbol\gamma}_1 + 
\bar{\mathbf{x}}_i^{\prime}\check{\boldsymbol\gamma}_2 = 0$, which is not possible if 
$\operatorname{plim}_{M\to\infty}\frac{1}{M}\sum_{m=1}^{M}\sum_{i=1}^{n_m}
\mathbf{z}_i\mathbf{z}_i^{\prime}$ is full rank. Hence, under the conditions given in Proposition \ref{PropoAri}, $\mathbf{k}_i$ cannot be linearly independent and Assumption \ref{H4}$(ii)$ holds.  \hfill $\blacksquare$
\newpage

\appendix
\begin{center}
    \Large \bf Online Appendix
\end{center}
\section{Heterogeneity, Non-Equivalence of the Spillover and Conformity Models and a Specification Test}\label{Appendix0}

In this Appendix, I show that the best response functions from the spillover and conformism models are not identical when the model accounts for action-specific heterogeneity. This allows me to propose specification tests to infer if each microfoundations is consistent with the data.

In contrast to the heterogeneous case, the spillover and conformity microfoundations are equivalent under homogeneous peer effects, differing only by a constant intercept shift. This equivalence was demonstrated by \cite{brock_discrete_2001} for $y_i=\{-1,1\}$. I first provide a brief exposition when $y_i=\{0,1\}$ and peer effects are homogeneous. To my knowledge, this derivation, although trivial, does not appear in the literature.  \\ 

\textbf{A.1 Spillover and conformity model with homogeneous peer effects} \\

\noindent Let $U^{spill}(y_i,\mathbf{y}_{-i})$ and $U^{conf}(y_i,\mathbf{y}_{-i})$ be the utility functions in the spillover and conformity models, respectively, and given by:
$$U^{spill}(y_i,\mathbf{y}_{-i})=\alpha_iy_i+\beta^{spill}\bar{y}_iy_i+\beta^{spill}(1-\bar{y}_i)(1-y_i)+\epsilon(y_i)$$
and 
$$U^{conf}(y_i,\mathbf{y}_{-i})=\alpha_i y_i-\frac{\beta^{conf}}{2}\left(y_i-\bar{y}_i\right)^2+\epsilon(y_i)$$
where $U^{spill}(\cdot)$ is similar to the utility function in \cite{xu2018}. The utility of a player who selects the high (low) action increases with the share of her friends who play the high (low) action. Since $y_i$ is binary, this model also implies that the utility of a player who selects the high (low) action decreases with the share of her friends who play the low (high) action.
Under Assumptions \ref{H1} and \ref{H2}, and the same derivation as in the main text, the best responses are given by:

$$\text{Pr}^{spill}(y_i=1\mid \boldsymbol{\alpha},\mathbf{G})=F_\eta\left(\alpha_i+2\beta^{spill}(\bar{p}_i-1)\right)$$
and 
$$\text{Pr}^{conf}(y_i=1\mid \boldsymbol{\alpha},\mathbf{G})=F_\eta\left(\alpha_i+\beta^{conf}\left(\bar{p}_i-\frac{1}{2}\right)\right)$$
 The best responses from the spillover and conformity models are then identical when peer effects are homogeneous, with $\beta^{conf}=2\beta^{spill}$. Hence, the researcher cannot identify the true microfoundation from the data. \\

\textbf{A.2 Spillover and conformity model with action-specific peer effects} \\

\noindent I now introduce action-specific heterogeneity. Observe that I formally account for isolated agents here, as their existence is necessary to derive the specification test in the next section. Let 
$$U^{spill}(y_i,\mathbf{y}_{-i})=\alpha_iy_i+\mathds{1}\{d_i>0\}\left[\beta^h\bar{y}_iy_i+\beta^l(1-\bar{y}_i)(1-y_i)+\epsilon(y_i)\right]$$
and 
$$U^{conf}(y_i,\mathbf{y}_{-i})=\alpha_i y_i-\mathds{1}\{d_i>0\}\left[\frac{\beta^h}{2}(y_i-\bar{y}_i)^2y_i+\frac{\beta^l}{2}(y_i-\bar{y}_i)^2(1-y_i)\right]+\epsilon(y_i)$$
where $U^{conf}(\cdot)$ corresponds to Equation \eqref{U}, the utility function of the heterogeneous conformity model studied in the main text. 

Under Assumptions \ref{H1} and \ref{H2}, and the same derivation as in the main text, the best responses are given by:
\begin{equation}\label{A.1}
    \text{Pr}^{spill}(y_i=1\mid \boldsymbol{\alpha},\mathbf{G})=F_\eta\left(\alpha_i+(\beta^h+\beta^l)\bar{p}_i-\mathds{1}\{d_i>0\}\beta^l\right) \tag{A.1}
\end{equation}

and 
\begin{equation}\label{A.2}
  \text{Pr}^{conf}(y_i=1\mid \boldsymbol{\alpha},\mathbf{G})=F_\eta\left(\alpha_i+\mathds{1}\{d_i>0\}\left[\beta^h\left(\bar{p}_i-\frac{1}{2}\right)+\frac{1}{2}\Delta \beta \left(\mathbf{g}_i\boldsymbol\Sigma  \mathbf{g}^{\prime}_i\right)\right]\right) \tag{A.2}
\end{equation}
where the last equation corresponds to Equation \eqref{eq7}. The best response functions are not identical unless $\beta^h=\beta^l$, i.e., if the taste for conformity is homogeneous across actions. \\

\textbf{A.3 Specification test} \\

\noindent I define the following \textit{generalized} reduced-form equation:

\begin{equation}\label{gen}
\Pr(y_i=1 \mid \boldsymbol{\alpha}, \mathbf{G}) = F_\eta \Bigg(
\alpha_i + \beta_1 \bar{p}_i + \beta_2 \mathds{1}\{d_i > 0\} + \beta_3 (\mathbf{g}_i \boldsymbol{\Sigma} \mathbf{g}_i') 
\Bigg). \tag{A.3}
\end{equation}

\noindent This generalized reduced-form equation nests \textit{both} best response functions from the spillover model (Equation \eqref{A.1}) and the conformity model (Equation \eqref{A.2}). Specifically, the mappings are as follows:

\begin{itemize}
    \item Spillover model: $\beta_1 = \beta^l + \beta^h$, $\beta_2 = -\beta^l$, $\beta_3 = 0$.
    \item Conformity model: $\beta_1 = \beta^h$, $\beta_2 = -\frac{\beta^h}{2}$, $\beta_3 = \frac{\beta^l - \beta^h}{2}$.
\end{itemize}
Observe that the identification and estimation of Equation \eqref{gen} are similar to those of Equation \eqref{eq7}. The only additional assumption required is that the dummy variable indicating if a player is isolated or not, $\mathds{1}\{d_i > 0\}$, exhibits variation across players, i.e., there must be isolated and non-isolated agents in the at least one network.
\noindent It is then straightforward to define specification tests for each model:

\begin{itemize}
    \item Spillover model: The null hypothesis is $H_0: \beta_3 = 0$, which implies that a pure spillover model is consistent with the data. The alternative hypothesis is $H_1: \beta_3 \neq 0$ and implies that a pure spillover model is not consistent with the data. This restriction can be tested using a Wald test.
    
    \item Conformity model: The null hypothesis is $H_0: \beta_1 = -2\beta_2$, which implies that a pure conformity model is consistent with the data. The alternative hypothesis is $H_1: \beta_1 \neq -2\beta_2$, and a Wald test can also be applied here.
\end{itemize}

\newpage
\section{Alternative Social Distance Functions and Outcome Coding} \label{AppendixA}

In this Appendix, I present several alternative social distance functions and show that action-specific heterogeneity in the preference for conformity can also be identified in the resulting conformity models. I consider first an \textit{aggregate} social distance function that computes the average (or sum, à la \cite{Boucher2016}) of distances from friends' actions rather than the deviation from the average action of friends. Then, I study a linear social distance function, such as the one used by \cite{akerlof1980}, \cite{kandel1992} or \cite{fershtman1998}. However, identification in both models relies on the existence of isolated players. Since, in some network data, none of the agents are isolated, I focus in the main text on the quadratic distance function.

In addition, I show that using the alternative coding $y_i=\{-1,1\}$, as in \cite{lee_binary_2014} and \cite{brock_discrete_2001}, results in an identification failure if action-specific heterogeneity is introduced.\\

%\footnote{The local-aggregate social distance function introduced by \cite{Boucher2016}, $\widetilde{S}(y_i,p_{-i})=\frac{\beta}{2}A\left[\left(y_i-p\right)^2\right]$, yields an identified asymmetric peer effects model where players incur a cost proportional to the number of peers selecting a different action rather than proportional to the distance from the average action of their peers. The equilibrium choice probabilities are given by $p^*=F_\eta\left(\alpha_i+\mathds{1}\{d>0\}\left\{\beta^h\left(Ap^*-\frac{1}{2}\right)+\frac{1}{2}\Delta\beta(Ap^*)^2\right\}\right)$, which is identified under conditions similar to those stated in Assumption \ref{H4}.}. \\

\textbf{B.1. A social distance function that computes the weighted sum of distances from friends' actions rather than the distance to the weighted sum of friends' actions} \\

\noindent I use the term \textit{weighted sum} so that the social distance introduced here accommodates both the average (when the interaction matrix is row-normalized and the weights are $\frac{1}{d_i}$) and the sum (when the interaction matrix is the adjacency matrix). The interaction matrix is denoted $\mathbf{W}=[w_{ij}]$. Here also, I account for isolated players as their existence is necessary to identify the parameters. 
Consider the following heterogeneous social distance function, which computes the weighted sum of the distance between the player's action and the actions of her friends, rather than the distance between the player's action and the weighted sum of her friends' actions:

$$\Tilde{S}(y_i,\mathbf{y}_{-i})=y_i\frac{\beta^h}{2}\sum_{j\neq i}w_{ij}\left(y_i-y_j\right)^2+ (1-y_i)\frac{\beta^l}{2}\sum_{j\neq i}w_{ij}\left(y_i-y_j\right)^2$$
When $\mathbf{W}=\mathbf{A}$, this distance function aggregates the deviations of a player from all of her friends’ actions and scales with the players’ degrees. This specification may be empirically relevant, as it captures the idea that the pressure to conform could be stronger (or weaker) in larger groups. 
Recognizing that $y_i=1 \implies y_i\geq y_j$ and $y_i=0 \implies y_i\leq y_j$, the expected utility function derived from $\Tilde{S}$ can be written as:
\begin{align*}
   \mathbb{E}\mid U_i(y_i,\mathbf{y}_{-i})] &=\begin{cases} 
\alpha_i-\mathds{1}\left\{d_i>0\right\} \mathbb{E}\left[\Tilde{S}(1,\mathbf{y}_{-i})\mid \boldsymbol \alpha,\mathbf W\right]+\epsilon_i(1) & \text{if } y_i=1 \\
-\mathds{1}\left\{d_i>0\right\} \mathbb{E}\left[\Tilde{S}(0,\mathbf{y}_{-i})\mid \boldsymbol \alpha,\mathbf W\right]+\epsilon_i(0) & \text{if } y_i=0 \end{cases}\\
   &=\begin{cases} 
\alpha_i-\mathds{1}\left\{d_i>0\right\}\frac{\beta^h}{2}\sum_{j \neq i}w_{ij}\left(1-2\mathbb{E}\left[y_j\mid \boldsymbol \alpha,\mathbf W\right]+\mathbb{E}\left[y_j^2\mid \boldsymbol \alpha,\mathbf W\right]\right)+\epsilon_i(1) & \text{if } y_i=1 \\
-\mathds{1}\left\{d_i>0\right\}  \frac{\beta^l}{2}\sum_{j \neq i}w_{ij}\left(\mathbb{E}\left[y_j^2\mid \boldsymbol \alpha,\mathbf W\right]\right)+\epsilon_i(0) & \text{if } y_i=0 \\
\end{cases} \\
%&=\begin{cases} 
%\alpha_i-\mathds{1}\left(d_i>0\right)\frac{\beta^h}{2}\sum_{j \neq i}g_{ij}\left(1-2p_j +p_j\right)+\epsilon_i(1) & \text{if } y_i=1 \\
%-\mathds{1}\left(d_i>0\right)  \frac{\beta^l}{2}\sum_{j \neq i}g_{ij}p_j+\epsilon_i(0) & \text{if } y_i=0 \\
%\end{cases} \\
&=\begin{cases} 
\alpha_i-\mathds{1}\left\{d_i>0\right\}\frac{\beta^h}{2}\left(1-\mathbf{w}_i\mathbf{p}\right)+\epsilon_i(1) & \text{if } y_i=1 \\
-\mathds{1}\left\{d_i>0\right\}  \frac{\beta^l}{2}\mathbf{w}_i\mathbf{p}+\epsilon_i(0) & \text{if } y_i=0 \\
\end{cases} \\
\end{align*}
where $\mathbb{E}\left[y_j^2\mid \boldsymbol \alpha,\mathbf W\right]=\mathbb{E}\left[y_j\mid \boldsymbol \alpha,\mathbf W\right]$ since $y_j$ is a binary outcome.
Using the same derivation of the players' decision rule as in the paper, I obtain the following best response strategies:
\begin{equation}\label{A1}
   \mathbf{p}^*=F_{\eta}\left( \boldsymbol\alpha +\frac{\beta^h}{2}\left(\mathbf{W}\mathbf{p}^*-\mathds{1}\left\{\mathbf{d}>\mathbf{0}_n\right\}\right)+\frac{\beta^l}{2}\mathbf{W}\mathbf{p}^*\right) \tag{B.1}
\end{equation}
It is straightforward to use the approach given in Appendix \ref{AppendixB} to show that the equilibrium is unique if $\lvert \beta^h + \beta^l\rvert \leq \frac{2}{\max_u f_{ \eta}(u)}$.
However, the identification strategy differs from the model presented in the main text because $\beta^h$ and $\beta^l$ are separately identified only in the presence of isolated players in the network. The use of isolated players for identification has also been emphasized recently by \cite{boucher2023}. In the present context, the existence of isolated players ensures the identification of $\beta^h$. To see this, recall that $\alpha_i=m_i^{\prime}\boldsymbol{\gamma_0} + x_i^{\prime}\boldsymbol\gamma_1 +\bar{\mathbf{x}}_i^{\prime}\boldsymbol\gamma_2$ and note that $\bar{p}^*_i=0$ if $i$ is isolated or if none of her friends play the high action. Consider first the best-response of isolated players, denoted by the $iso$ index: $$\mathbf{p}_{iso}=F_{\eta}\left(\mathbf{m}_{iso}^{\prime}\boldsymbol{\gamma_0} + \mathbf{x}_{iso}^{\prime}\boldsymbol\gamma_1 +\bar{\mathbf{x}}_{iso}^{\prime}\boldsymbol\gamma_2\right)$$ 
Assuming, as in \cite{boucher2023}, that the coefficients $(\boldsymbol\gamma_0,\boldsymbol\gamma_1,\boldsymbol\gamma_2)$ are the same among the isolated and non-isolated sub-samples, these parameters are identified. Then, consider the non-isolated players who face a norm at 0, i.e. $\bar{p}_i=0$, indexed by $(niso,0)$:
$$\mathbf{p}_{niso,0}=F_{\eta}\left(\mathbf{m}_{niso,0}^{\prime}\boldsymbol{\gamma_0} + \mathbf{x}_{niso,0}^{\prime}\boldsymbol\gamma_1 +\bar{\mathbf{x}}_{niso,0}^{\prime}\boldsymbol\gamma_2-\frac{\beta^h}{2}\right)$$
Because $(\boldsymbol\gamma_0,\boldsymbol\gamma_1,\boldsymbol\gamma_2)$ is identified, $\beta^h$ can be identified. Finally, consider the non-isolated players who face a positive norm, i.e. $0<\bar{p}_i\leq 1$ and index them by $({niso,1})$:
$$\mathbf{p}_{niso,1}=F_{\eta}\left(\mathbf{m}_{niso,1}^{\prime}\boldsymbol{\gamma_0} + \mathbf{x}_{niso,1}^{\prime}\boldsymbol\gamma_1 +\bar{\mathbf{x}}_{niso,1}^{\prime}\boldsymbol\gamma_2+\frac{\beta^h}{2}\left(\bar{\mathbf{p}}_{niso,1}-\mathds{1}\left\{d_{niso,1}>0\right\}\right)+\frac{\beta^l}{2}\bar{\mathbf{p}}_{niso,1}\right)$$
Assuming that the taste for conformity when choosing the high action is identical between the nonisolated players that face a norm at zero and those that face a strictly positive one, $\beta^l$ is identified. \\

\textbf{B.2. Linear social distance function} \\

\noindent Next, consider an alternative linear social distance function such as the one used by \cite{akerlof1980}, \cite{kandel1992} or \cite{fershtman1998}:\footnote{The associated utility function also corresponds to the one used in the inequality aversion model of \cite{fehr1999}.}
\begin{equation*}
S(y_i,\mathbf{y}_{-i})=\beta\lvert y_i-\bar{y}_{-i}\rvert
\end{equation*}
This distance function differs from the quadratic one because small deviation from the norm are as penalized (relatively) compared to large ones, i.e., $\frac{\partial S(y_i,\mathbf{y}_{-i})}{\partial (y_i-\bar{y}_{-i})} = \beta$, whereas the quadratic social distance function yields $\frac{\partial S^{quad}(y_i,\mathbf{y}_{-i})}{\partial (y_i-\bar{y}_{-i})} = 2\beta\lvert y_i-\bar{y}_{-i}\rvert $. 
The linear distance function with action-specific heterogeneity can be written as: 
\begin{equation*}
S^{lin}(y_i,\mathbf{y}_{-i})=y_i\beta^h(y_i-\bar{y}_{-i})+(1-y_i)\beta^l(\bar{y}_{-i}-y_i)
\end{equation*}
and the associated expected utility is given by:
\begin{align*}
 \mathbb{E}[U_i(y_i,\mathbf{y}_{-i})]&= \begin{cases} 
\alpha_i-\mathds{1}\left\{d_i>0\right\} \beta^h\sum_{j\neq i}g_{ij}(1-p_j)+\epsilon_i(1) & \text{if } y_i=1 \\
-\mathds{1}\left\{d_i>0\right\}\beta^l\sum_{j\neq i}g_{ij}p_j+\epsilon_i(0) & \text{if } y_i=0 \\
\end{cases}\\
&= \begin{cases} 
\alpha_i+\beta^h\left(\bar{p}_i-\mathds{1}\left\{d_i>0\right\}\right)+\epsilon_i(1) & \text{if } y_i=1 \\
-\beta^l\bar{p}_i+\epsilon_i(0) & \text{if } y_i=0 \\
\end{cases}
\end{align*}
Which yields the following best responses at the BNE:
\begin{equation}\label{A2}
\mathbf p^*=F_{\eta}\left(\alpha+\beta^h\left(\bar{\mathbf{p}}^*-\mathds{1}\left\{\mathbf{d}>\mathbf{0}_n\right\}\right)+\beta^l\bar{\mathbf{p}}^*\right) \tag{B.2}
\end{equation}
Note that the best response function with the linear distance function given in Equation \eqref{A2} is a rescaled version of the best response function derived from the \textit{aggregate} quadratic distance function, given in Equation \eqref{A1}. This implies that the equilibrium is unique in the linear conformity model if the conformity parameters satisfy $ \lvert \beta^h+\beta^l\rvert < \frac{1}{\max_u f_{ \eta}(u)}$ and that the identification strategy via isolated players can also be applied.\\

%In addition, observe that Equations \eqref{A1} and \eqref{A2} can be obtained as special cases of the \textit{generalized} reduced-form equation given by Equation \eqref{gen}, with $\beta_1 = \frac{\beta^l + \beta^h}{2}$, $\beta_2 = -\frac{\beta^h}{2}$, $\beta_3 = 0$ and $\beta_1 = \beta^l + \beta^h$, $\beta_2 = -\beta^h$, $\beta_3 = 0$, respectively. Thus, a specification test can also be used to infer whether a conformity model with a linear or \textit{aggregate} social distance function is consistent with the data.  \\

\textbf{B.3. Alternative coding scheme} \\

\noindent Using the heterogeneous social distance function developed in this paper with \cite{brock_discrete_2001} and \cite{lee_binary_2014}'s coding scheme, $\mathcal{Y}_i=\left\{-1,1\right\}$, yields:
\begin{equation*}
\mathbf p^*=F_{ \eta}\left(2\alpha+\mathds{1}\left\{\mathbf{d}>\mathbf{0}_n\right\}\beta^h\left(\bar{\mathbf{p}}^*-\frac{1}{2}\mathbf{1}_n\right)+\mathds{1}\left\{\mathbf{d}>\mathbf{0}_n\right\}\beta^l\left(\bar{\mathbf{p}}^*+\frac{1}{2}\mathbf{1}_n\right)+ \frac{1}{2}\Delta\beta\text{diag}\left(\mathbf{G}\boldsymbol\Sigma \mathbf{G}^{\prime}\right)\right)
\end{equation*}
\noindent This model is not identified unless $\beta^h=\beta^l$, which corresponds to a homogeneous distance function and yields $p^*=F_\eta\left(2\alpha_i+2\beta \bar{\mathbf{p}}^*\right)$, i.e., the model proposed by \cite{brock_discrete_2001}.\footnote{This nonidentification result also holds with the social distance functions introduced in B.1 and B.2, $\Tilde{S}$ and $S^{lin}$, when $\mathcal{Y}_i=\left\{-1,1\right\}$, but is omitted for brevity.} Contrary to the models discussed in B.1 and B.2, identification fails here because non-isolated players that face a norm at zero do not help identify $\beta^h$ or $\beta^l$. Indeed, their best response function is $\mathbf{p}_{niso,0}=F_{\eta}\left(2\alpha_{niso,0}-\beta^h+\beta^l\right)$.

\clearpage
\newpage
\end{document}